\documentclass[a4paper,fleqn,usenatbib]{mnras}

\usepackage{newtxtext,newtxmath}

\usepackage[T1]{fontenc}
\usepackage{ae,aecompl}

\usepackage{graphicx}	
\newsavebox\CBox
\def\textBF#1{\sbox\CBox{#1}\resizebox{\wd\CBox}{\ht\CBox}{\textbf{#1}}}
\usepackage{amsmath}	
\usepackage{amssymb}	

\usepackage{amsfonts}
\usepackage{bm}
\usepackage{hyperref}
\usepackage{threeparttable}

\title[SMA imaging of bright S2CLS sources]{High-resolution SMA imaging of bright submillimetre sources from the SCUBA-2 Cosmology Legacy Survey}

\author[R. Hill et al.]{
Ryley Hill,\hyperlink{one}{$^{1}$}
Scott C.~Chapman,\hyperlink{two}{$^{2}$}
Douglas Scott,\hyperlink{one}{$^{1}$}
Glen Petitpas,\hyperlink{three}{$^{3}$}
Ian Smail,\hyperlink{four}{$^{4}$}\newauthor
Edward L.~Chapin,\hyperlink{five}{$^{5}$}
Mark A.~Gurwell,\hyperlink{three}{$^{3}$}
Ryan Perry,\hyperlink{two}{$^{2}$}
Andrew W.~Blain,\hyperlink{six}{$^{6}$}\newauthor
Malcolm N.~Bremer,\hyperlink{seven}{$^{7}$}
Chian-Chou Chen,\hyperlink{four}{$^{4}$}$^{, }$\hyperlink{eight}{$^{8}$}
James S.~Dunlop,\hyperlink{nine}{$^{9}$}
Duncan Farrah,\hyperlink{ten}{$^{10}$}\newauthor
Giovanni G.~Fazio,\hyperlink{three}{$^{3}$}
James E.~Geach,\hyperlink{eleven}{$^{11}$}
Paul Howson,\hyperlink{two}{$^{2}$}
R.~J.~Ivison,\hyperlink{eight}{$^{8}$}$^{, }$\hyperlink{nine}{$^{9}$}
Kevin Lacaille,\hyperlink{two}{$^{2}$}$^{, }$\hyperlink{twelve}{$^{12}$}\newauthor
Micha{\l} J.~Micha{\l}owski,\hyperlink{thirteen}{$^{13}$}
James M.~Simpson,\hyperlink{nine}{$^{9}$}$^{, }$\hyperlink{fourteen}{$^{14}$}
A.~M.~Swinbank,\hyperlink{four}{$^{4}$}\newauthor
Paul P.~van der Werf,\hyperlink{fifteen}{$^{15}$}
David J.~Wilner\hyperlink{three}{$^{3}$}
\\
\hypertarget{one}{$^{1}$}Department of Physics and Astronomy, University of British Columbia, 6225 Agricultural Road, Vancouver, V6T 1Z1, Canada\\
\hypertarget{two}{$^{2}$}Department of Physics and Atmospheric Science, Dalhousie University, 6310 Coburg Road, Halifax, B3H 4R2, Canada\\
\hypertarget{three}{$^{3}$}Harvard-Smithsonian Center for Astrophysics, 60 Garden Street, Cambridge, MA 02138, USA\\
\hypertarget{four}{$^{4}$}Centre for Extragalactic Astronomy, Department of Physics, Durham University, South Road,  Durham, DH1 3LE, UK\\
\hypertarget{five}{$^{5}$}Herzberg Astronomy and Astrophysics, National Research Council Canada, 5071 West Saanich Road, Victoria, V9E 2E7, Canada\\
\hypertarget{six}{$^{6}$}Department of Physics and Astronomy, University of Leicester, University Road, Leicester, LE1 7RH, UK\\
\hypertarget{seven}{$^{7}$}H.H. Wills Physics Laboratory, University of Bristol, Tyndall Avenue, Bristol, BS8 1TL, UK\\
\hypertarget{eight}{$^{8}$}European Southern Observatory, Karl-Schwarzschild-Stra{\ss}e 2, Garching, 85748, Germany\\
\hypertarget{nine}{$^{9}$}Institute for Astronomy, University of Edinburgh, Royal Observatory, Edinburgh, EH9 3HJ, UK\\
\hypertarget{ten}{$^{10}$}Department of Physics MC 0435, Virginia Polytechnic Institute and State University, 850 West Campus Drive, Blacksburg, VA 24061, USA\\
\hypertarget{eleven}{$^{11}$}Centre for Astrophysics Research, School of Physics, Astronomy and Mathematics, University of Hertfordshire, Roehyde Way, Hatfield, \\
\phantom{$^{11}$}AL10 9AB, UK\\
\hypertarget{twelve}{$^{12}$}Department of Physics and Astronomy, McMaster University, 1280 Main Street West, Hamilton, L8S 4M1, Canada\\
\hypertarget{thirteen}{$^{13}$}Astronomical Observatory Institute, Faculty of Physics, Adam Mickiewicz University, ul.~S{\l}oneczna 36, 60-286 Pozna{\'n}, Poland\\
\hypertarget{fourteen}{$^{14}$}Academia Sinica Institute of Astronomy and Astrophysics, No. 1, Sec. 4, Roosevelt Road, Taipei 10617, Taiwan\\
\hypertarget{fifteen}{$^{15}$}Leiden Observatory, Leiden University, P.O. box 9513, Leiden, NL-2300 RA, the Netherlands
}

\date{05 October 2017}

\pubyear{2017}

\begin{document}
\label{firstpage}
\pagerange{\pageref{firstpage}--\pageref{lastpage}}
\maketitle

\begin{abstract}
We have used the Submillimeter Array (SMA) at 860\,$\mu$m to observe the brightest sources in the SCUBA-2 Cosmology Legacy Survey (S2CLS). The goal of this survey is to exploit the large field of the S2CLS along with the resolution and sensitivity of the SMA to construct a large sample of these rare sources and to study their statistical properties. We have targeted 70 of the brightest single-dish SCUBA-2 850\,$\mu$m sources down to $S_{850}\,{\approx}\,8$\,mJy, achieving an average synthesized beam of 2.4\,arcsec and an average rms of $\sigma_{860}\,{=}\,1.5$\,mJy\,beam$^{-1}$ in our primary beam-corrected maps. We searched our SMA maps for $4\sigma$ peaks, corresponding to $S_{860}\,{\gtrsim}\,6$\,mJy sources, and detected 62, galaxies, including three pairs. We include in our study 35 archival observations, bringing our sample size to 105 bright single-dish submillimetre sources with interferometric follow-up. We compute the cumulative and differential number counts, finding them to overlap with previous single-dish survey number counts within the uncertainties, although our cumulative number count is systematically lower than the parent S2CLS cumulative number count by $14\,{\pm}\,6$ per cent between 11 and 15\,mJy. We estimate the probability that a ${\gtrsim}\,10$\,mJy single-dish submillimetre source resolves into two or more galaxies with similar flux densities to be less than 15 per cent. Assuming the remaining 85 per cent of the targets are ultra-luminous starburst galaxies between $z\,{=}\,2$--3, we find a likely volume density of ${\gtrsim}\,400$\,M$_{\sun}$\,yr$^{-1}$ sources to be ${\sim}\,3^{+0.7}_{-0.6}\,{\times}\,10^{-7}$\,Mpc$^{-3}$. We show that the descendants of these galaxies could be ${\gtrsim}\,4\,{\times}\,10^{11}$\,M$_{\sun}$ local quiescent galaxies, and that about 10 per cent of their total stellar mass would have formed during these short bursts of star-formation.
\end{abstract}




\section{Introduction}

The emergence of submillimetre (submm) astronomy has led to the discovery of a cosmologically important population of submm galaxies (SMGs), which appear to be among the earliest and most actively star-forming galaxies in the Universe, often reaching luminosities of a few times 10$^{13}$\,L$_{\sun}$ and star-formation rates (SFRs) greater than a few hundred M$_{\sun}$\,yr$^{-1}$ \citep[e.g.,][]{blain2002,magnelli2012,swinbank2014,mackenzie2017,michalowski2017} and above \citep[e.g. HFLS3, see][]{riechers2013} around redshifts 2--3 \citep[e.g.,][]{chapman2005,simpson2014,simpson2017}. The Submillimeter Common User Bolometer Array \citep[SCUBA;][]{holland1999}, mounted on the 15-m James Clerk Maxwell Telescope (JCMT), was the first multi-pixel instrument to detect this population of high-redshift SMGs \citep[e.g.][]{smail1997,barger1998,hughes1998}. This motivated the development of more sensitive detectors such as the second generation SCUBA-2 \citep{holland2013}, the Large Apex BOlometer CAmera \cite[LABOCA;][]{sirigno2009}, and the AZtronomical Thermal Emission Camera \citep[AzTEC;][]{wilson2008}, as well as the Balloon-borne Large Aperture Submillimeter Telescope \citep[BLAST;][]{pascale2008} and the space-based Spectral and Photometric Imaging REceiver \citep[SPIRE;][]{griffin2010} on board the {\it Herschel\/} satellite, all of which have been used to further investigate SMGs.

While single dish observations of SMGs were able to greatly increase our knowledge about the evolution of star formation in the Universe \citep[e.g.,][]{blain1999,magnelli2013,gruppioni2013,swinbank2014,koprowski2017}, their connection with today's galaxies remains unclear, although evidence is mounting that they are progenitors of massive elliptical galaxies \citep[e.g.,][]{lilly1999,scott2002,genzel2003,swinbank2006,toft2014,simpson2014,koprowski2014,vandokkum2015,koprowski2016,michalowski2017,simpson2017}. There is also debate about whether or not mergers are important for SMGs. Many simulations require mergers to achieve the observed massive SFRs \citep[e.g.,][]{narayanan2015} while others do not \citep[e.g.,][]{dave2010}, and on the other hand,
observations of physically associated pairs of SMGs with disturbed gas morphologies indicate that mergers are present \citep[e.g.,][]{tacconi2008,engel2010,chen2015}, while ultra-luminous SMGs have been seen that lack evidence of multiplicity and fit on the high-mass end of the `main sequence' of star-forming galaxies \citep[e.g.,][]{targett2013,michalowski2017}. Progress is impeded by the sub-optimal angular resolution offered by single dish telescopes at submm wavelengths, which typically ranges between 10 and 30 arcseconds. At these scales, source blending becomes a significant problem, and optical/near-infrared (NIR) counterparts cannot be easily identified. 

This problem was first tackled by exploiting the high spatial resolution available to interferometers operating in the radio waveband, where synchrotron emission linked to supernovae is correlated with far-infrared (FIR) emission from dust \citep[e.g.,][]{condon1992,yun2001,ivison2010, magnelli2015} -- dust thought to be created following those same supernova events \citep[e.g.][]{indebetouw2014} and heated by young, massive stars.  Radio studies of SMGs were typically able to determine positions to subarcsecond accuracy, and thus localize multiwavelength counterpart galaxies using probabilistic arguments \citep[e.g.,][]{chapman2001,ivison2002,chapman2002,chapman2003,bertoldi2007,biggs2011}, which greatly improved our understanding of their redshift distribution \citep{smail2000,chapman2003,chapman2005,dannerbauer2004,smolcic2012b} and physical characteristics \citep{ivison1998,ivison2000,smail2000,chapman2004,borys2004}. In particular, \citet{ivison2007} showed that a significantly larger fraction of SMGs contained multiple radio counterparts than would be expected by chance, suggesting therefore that they could comprise groups of physically associated galaxies.

However, more accurately pinpointing the submm emission directly -- the only way to be fully sure that the associated positions and optical/infrared (IR) counterparts are {\it bone fide} -- was not possible until the leap in continuum sensitivity provided by new submm interferometers and wide-bandwidth correlators, such as those available at the Plateau de Bure Interferometer \citep[PdBI;][]{guilloteau1992}, the Submillimeter Array \citep[SMA;][]{ho2004} and, most recently, the Atacama Large Millimeter/submillimeter Array \citep[ALMA;][]{wootten2009}. These have greatly aided the localisation of counterparts and the further characterization of SMGs.  These facilities were able to confirm that many SMGs exhibit multiplicity (e.g., \citealt{iono2006}; \citealt{younger2007}; \citealt{younger2009}; \citealt{wang2011}; \citealt{smolcic2012}; \citealt{hodge2013}; \citealt{simpson2015}; \citealt{miettinen2015}; Stach et al.~in preparation), where one bright single-dish submm source resolves into two or three individual SMGs.

Large single-dish submm surveys \citep[e.g.,][]{scott2002,greve2004,wang2004,coppin2006,bertoldi2007,weiss2009,oliver2010,valiante2016,geach2016}, followed up by interferometers, have been important for identifying large numbers of SMGs for multiwavelength follow-up as they provide substantial catalogues of bright single-dish sources across continuous patches of sky that interferometers can follow-up. For example, \citet{barger2012} used the SMA at 870\,$\mu$m to observe 16 $S_{850}\,{>}\,3$\,mJy sources detected with SCUBA-2 in the Great Observatories Origins Deep Survey-North field \citep[GOODS-N;][]{wang2004}, and similarly \citet{smolcic2012} used the PdBI at 1.3\,mm to target 28 $S_{870}\,{>}\,5$\,mJy sources detected by LABOCA at 870\,$\mu$m in the COSMOS field (Navarette et al.~in preparation). A larger LABOCA 0.25\,deg$^{2}$ survey of the Extended {\it Chandra\/} Deep Field South \citep[LESS;][]{weiss2009} was followed up with ALMA by \citet{hodge2013}, who observed 126 sources $S_{870}\,{>}\,3.5$\,mJy, and more recently, \citet{simpson2015} used ALMA at 870\,$\mu$m to follow-up 30 of the brightest ($S_{850}\,{>}\,5$\,mJy) sources detected in the UKIDSS-UDS field at 850\,$\mu$m, mapped by SCUBA-2 as part of the SCUBA-2 Cosmology Legacy Survey \citep[S2CLS;][]{geach2016}. While these types of surveys have begun to reach statistically significant numbers of samples, they nonetheless lack large numbers of the brightest single dish detected sources; for example, the LESS survey contained 20 sources with $S_{850}\,{>}\,8$\,mJy and six sources with $S_{850}\,{>}\,10$\,mJy, and the observations from \citet{simpson2015} contained 13 sources with $S_{850}\,{>}\,8$\,mJy and seven sources with $S_{850}\,{>}\,10$\,mJy when considering the final S2CLS catalogue. 

To date, the largest submm survey of the extragalactic sky is the complete S2CLS, encompassing 5\,deg$^{2}$ of the sky over seven cosmological fields: UKIDSS-UDS, COSMOS, {\it Akari\/}-NEP, Extended Groth Strip, Lockman Hole North, SSA22 and GOODS-North. The S2CLS detected over 2800 submm sources above $3.5\sigma$, where 114 of them had $S_{850}\,{>}\,8$\,mJy and 46 of them had $S_{850}\,{>}\,10$\,mJy. This survey is therefore well-suited to study the properties the brightest SMGs known to exist, thus clarifying such issues as the importance of mergers in galaxy formation and probing the highest ends of the luminosity and mass functions. In addition, SCUBA-2's location in the northern hemisphere makes northern fields such as the Extended Groth Strip and the Lockman Hole North distinctly observable with the SMA, thus providing a unique data set.

In this paper we present results from the largest yet interferometric follow-up programme of the brightest submm sources. We have imaged 70 SCUBA-2 sources with $S_{850}\,{\gtrsim}\,8$\,mJy at 2.4\,arcsec resolution using the SMA at 860\,$\mu$m, selected from 80 per cent of the available are in the S2CLS. In Section \ref{observations} we describe our target selection, data reduction and source extraction procedure, in Section \ref{analysis} we correct our flux density measurements for flux boosting and compare our data to the S2CLS catalogue to asses the reliability of our sample, and in Section \ref{discussion} we examine the completeness of our sample, present number counts, discuss the effects of multiplicity and investigate some properties of the population of bright SMGs seen in our data. We give our conclusions in Section \ref{conclusion}.

\section{Observations and data reduction}\label{observations}

\subsection{Target selection}

In our observing programme we used the SMA in the compact configuration at 860\,$\mu$m to investigate bright sources in five out of the seven wide 850-$\mu$m S2CLS fields, namely UKIDSS-UDS, COSMOS, the Extended Groth Strip, the Lockman Hole North, and SSA22 (hereafter the UDS, COSMOS, EGS, LHN and SSA22 fields, respectively). Combined, these fields make up about 80 per cent of the full S2CLS at 4\,deg$^{2}$, and contain 98 sources with $S_{850}\,{>}\,8$\,mJy and 39 sources with $S_{850}\,{>}\,10$\,mJy. Our initial aim was to target and resolve all sources down to ${\approx}\,8$\,mJy. At the time these observations were first proposed, the S2CLS had not yet been completed, being at that point shallower than the final maps published in \citet{geach2016}. This led to several cases where either a proposed SCUBA-2 target ended up fainter than expected, or an originally faint SCUBA-2 source ended up being brighter than 10\,mJy at 850\,$\mu$m. When selecting targets we only considered the measured (uncorrected) SCUBA-2 flux densities, which are believed to be boosted by positive noise and faint background galaxies that on average add a positive bias to the flux densities and are statistically corrected for in the final S2CLS catalogue in \citet{geach2016}. This effect resulted in more examples of apparently bright SCUBA-2 sources ending up being fainter in the final list.

There are several submm interferometric data sets in the literature that we did not re-observe in our programme. In the COSMOS field, \citet{younger2007,younger2009} selected the 15 highest significance sources in an AzTEC 1.1\,mm survey \citep{scott2008} for follow-up with the SMA at 890\,$\mu$m, and  later \citet{aravena2010} used the SMA at 890\,$\mu$m to observe two of the most significant sources detected in a Max Planck Millimeter Bolometer (MAMBO) 1.2\,mm survey \citep{bertoldi2007}. Later, \citet{smolcic2012b} followed up three more bright MAMBO- and AzTEC-selected sources \citep{aretxaga2011} with the Combined Array for Research in Millimeter-wave Astronomy (CARMA) at 1.3\,mm, and then \citet{smolcic2012} followed up another 28 sources detected by a LABOCA 870-$\mu$m survey (Navarette et al.~in preparation) with the PdBI at 1.3\,mm. In the UDS field, \citet{simpson2015} carried out a follow-up campaign of 30 bright S2CLS sources with ALMA at 870\,$\mu$m, and in the SSA22 field, \citet{tamura2010} used the SMA at 860\,$\mu$m to follow up the brightest source detected in a 1.1\,mm AzTEC survey \citep{tamura2009}. Additionally, there is a single strong gravitational lens in the UDS field, dubbed `Orochi', reaching an 850-$\mu$m flux density of 52.7\,mJy in the SCUBA-2 map; this source was followed up by \cite{ikarashi2011} with the SMA at 860\,$\mu$m in part of a detailed multiwavelength study. We have included 35 observations from these works into our analysis, and we describe these sources in further detail in Section \ref{completeness}.

\begin{figure}
\includegraphics[width=\columnwidth]{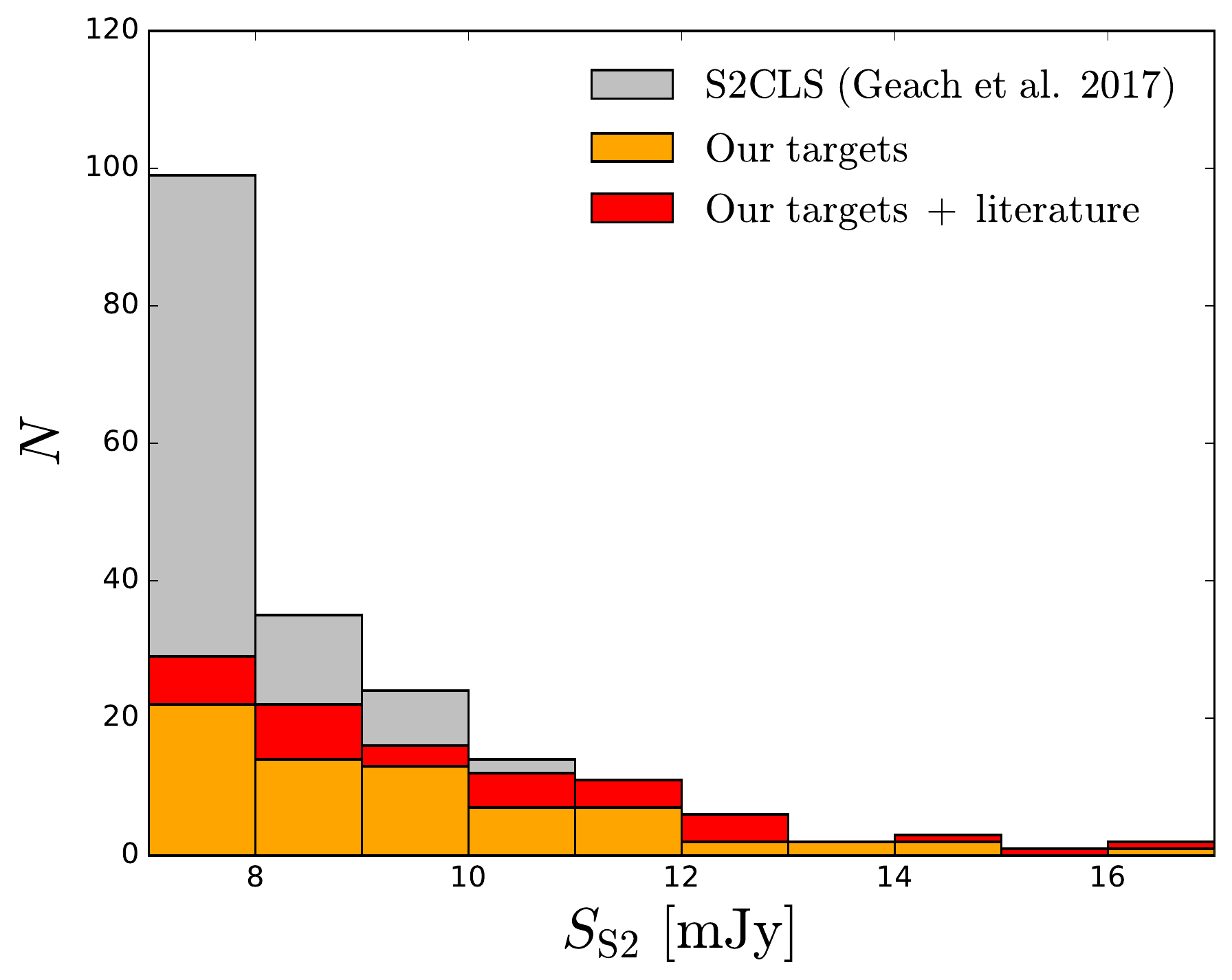}
\caption{Histogram showing the deboosted flux density distribution of the parent S2CLS survey from \citet{geach2016}, our 70 targets, and our full catalogue including these 70 targets and 35 archival sources from \citet{younger2007,younger2009}, \citet{aravena2010}, \citet{tamura2010}, \citet{ikarashi2011}, \citet{smolcic2012b,smolcic2012} and \citet{simpson2015}, which are included in our counts analysis. Our sample is a nearly complete selection of single-dish 850-$\mu$m SCUBA-2 sources with flux densities brighter than 10\,mJy.}
\label{sample}
\end{figure}

Our final SMA follow-up campaign sample consisted of 70 total targets; 23 in the UDS field, eight in the SSA22 field, 12 in the COSMOS field, 18 in the LHN field and nine in the EGS field. These sources had the brightest 850\,$\mu$m flux densities down to approximately 10\,mJy, except in the UDS field where we probed sources with flux densities down to about 8\,mJy. In Fig.~\ref{sample} we show the SCUBA-2 deboosted flux density distribution from our parent S2CLS sample, with the distribution of our targets and the distribution of our full catalogue (including archival sources) overlaid. This shows the completeness of our selection, which we quantify later in Section \ref{completeness}.

We note that we followed up two additional sources in the EGS field (EGS07 and EGS09) as well as two additional sources in the COSMOS field (COSMOS01 and COSMOS02) that ended up excluded from the final S2CLS catalogue. These four sources lie near the edge of the EGS and COSMOS maps, where the root mean square (rms) is higher, and were thus excluded from the area used to define the final S2CLS regions. COSMOS02 was confirmed to be the brightest S2CLS source in a follow-up program to achieve deeper imaging in the COSMOS field (Simpson et al.~in preparation), so it's clear that there is some interest in these sources. While these four sources do not appear in our study, we nonetheless report them here for completeness.

\subsection{SMA observations}

We targeted 70 bright SCUBA-2 sources from the S2CLS fields using the SMA, carried out over a period of two years between November 2014 and November 2016. The sources are widely spaced on the sky and there was never an opportunity to have more than one source. We note that the FWHM of SCUBA-2, about 14.8\,arcsec at 850\,$\mu$m, is significantly smaller than the primary beamsize of the SMA, about 35\,arcsec at 860\,$\mu$m, thus it is highly unlikely that any submm counterpart sources would lie outside of our observations. We set up the SMA in the compact configuration tuned to 345\,GHz with a bandwidth of at least 4\,GHz and had available between six and eight 6-m dishes for a given track. The upgrade of the SMA with the SWARM correlator during this period led to a steady increase in bandwidth during the course of our observing program, culminating in the final track using the full 32\,GHz of SWARM. This upgrade considerably improved the continuum sensitivity and made calibrations with fainter quasars easier as the program went on. We adopted track sharing, typically three sources per track, to provide the best possible $uv$-plane coverage of each source (given the limited number of antennas available with the SMA), with some sources repeated on multiple tracks to achieve our desired sensitivity. The synthesized beam achieved in this set up was on average 2.4\,arcsec full-width at half-maximum (FWHM) with our natural weighting of the visibilities, but the beam shape in some cases ranged in elongation on the major axis to 3.5\,arcsec. 

We performed the calibration and data inspection using the  {\sc idl}-based Caltech package {\sc mir} modified for the SMA. We generated continuum data by averaging the spectral channels after doing the passband phase calibration. We used both gain calibrators to derive gain curves. For consistency checks, we compared these results with those obtained by adopting just one calibrator. We did not find any systematic differences. We computed the fluxes using calibrators observed on the same day and under similar conditions (time, hour angle, and elevation). Flux densities were calibrated using typically Uranus, Neptune, Callisto or Titan, depending on availability and proximity to the given target. The flux calibration error is typically within ${\sim}\,10\%$. Observations ranged in conditions but typically had a precipitable water vapor (PVW) significantly less than 2\,mm ($\tau_{\rm 225\,GHz}\,{<}\,0.12$). Our general goal was to detect 100 per cent of a target's SCUBA-2 flux density at $4\sigma$, however the resulting sensitivities on a given source were mainly determined by scheduling, weather, and available antennas on a given night.

\subsection{Source detections}

We exported the calibrated interferometric visibility data to the package \textsc{miriad} for subsequent imaging and analysis. We weighted the visibility data inversely proportional to the system temperature and Fourier transformed them to form images. We used natural weighting to maximize the signal-to-noise (S/N). We {\sc clean}ed the images around detected sources to approximately 1.5 times the noise level to remove the effects of sidelobes (the results were not sensitive to choosing a slightly deeper {\sc clean}ing level, such as 1.0 times the noise). We typically achieved an rms between 1 and 2\,mJy\,beam$^{-1}$, but occasionally we were substantially deeper than this with very good weather and with the correlator working well. We corrected the images for the SMA primary beam response. 

We set a detection threshold of ${>}\,4\sigma$ peaks in our maps. We measured source positions and flux densities by fitting the peaks in the dirty images (which is known to be more reliable for interferometric data lacking extensive uv coverage due to contamination from imaging artifacts) and also fitting the images with point-source models using the {\sc miriad} {\tt imfit} routine. The results of both approaches were very consistent, and we adopted the former for further analysis. All of the SMA flux densities and flux density errors that we quote henceforth are primary-beam corrected.

In the UDS field we detected 21 out of the 23 SCUBA-2 sources we followed-up; none of these 21 sources were seen to break up into two components, and two sources remained undetected. Of the eight SCUBA-2 sources targeted in the SSA22 field, four were not detected above the $4\sigma$ level in the SMA maps, and in the remaining four we found single galaxies. Within the COSMOS field, our SMA observations detected a total of 10 galaxies from the 12 SCUBA-2 sources: one source broke up into two galaxies; and in three sources we found no peaks greater than $4\sigma$. In the LHN field we found 18 galaxies from our targeted sample of 18 SCUBA-2 sources. Of these 18 detections two are SCUBA-2 sources that break up into two galaxies, and in two cases we did not find any galaxies. In the EGS field we have detected single galaxies for all nine SCUBA-2 sources. We also report detections of all four of the SCUBA-2 sources we followed up outside of the boundary of the S2CLS regions, and note that none resolved into multiples.

Overall we detected 62 submm galaxies in 70 SMA pointings above a $4\sigma$ depth of about 6\,mJy. These detections are summarized in Tables \ref{table1}--\ref{table5}, where we provide the positions of both the SCUBA-2 sources and our SMA detections, the measured and deboosted SCUBA-2 flux densities of each target as $S_{{\rm S2}}^{{\rm obs}}$ and $S_{{\rm S2}}$, respectively, and our measured flux densities as $S_{{\rm SMA}}^{{\rm obs}}$. For undetected sources, we report the $4\sigma$ flux density limit achieved by our observations instead. In each field, we sort sources in descending order of their deboosted SCUBA-2 flux density. In Fig.~\ref{cutouts} we show SMA contours for each observation overlaid over existing {\it Spitzer\/}-Infrared Array Camera (IRAC) 3.6$\mu$m and Very Large Array (VLA) 1.4\,GHz images, when available, and over theparent SCUBA-2 850$\mu$m images. We can see that there are IR/radio sources coincidental with nearly all of our SMA positions to within 1\,arcsec, yielding robustly identified counterpart galaxies. The multiwavelength properties of these galaxies will be investigated in future work.

It is worth noting that in the COSMOS field, out of the 18 SCUBA-2 sources found by \citet{michalowski2017} to have multiwavelength counterparts and included in our sample, all were confirmed by our SMA imaging. In the UDS field, out of 35 SCUBA-2 sources overlapping between our two studies, 31 were confirmed (89 per cent), consistent with the reliability of ${\simeq}\,92$ per cent measured by \citet{michalowski2017} based on the ALMA data of \citet{simpson2015}. Similarly, \citet{chen2015} was able to identify multiwavelength counterparts for ${\simeq}\,79$ per cent of the SCUBA-2 sources detected in the Extended {\it Chandra\/} Deep Field South, consistent with our observations.

\section{Analysis}\label{analysis}

\subsection{Flux boosting}

The effects of selection biases, particularly `flux boosting', on our results are complicated. This is because we picked bright outliers in large SCUBA-2 maps and followed them up with the SMA at higher resolution. Because of this complexity, we put considerable effort into simulating our observing and analysis procedure. The effect of flux boosting results from the statistical nature of measuring flux densities in a noisy map where there are many more faint sources than bright ones. This effect will tend to scatter sources to higher flux densities rather than lower ones, hence the term `boosting'. 

Our approach follows that outlined in \citet{coppin2005,coppin2006}. We construct a prior distribution for the underlying flux density of sources in our maps by performing a set of simulations that reconstruct our observing strategy as follows. For each of the five fields in our study we first produced a mock SCUBA-2 map by injecting sources into an area of blank sky matching the area surveyed in the S2CLS. The flux densities were drawn from a Schechter-type function of the form

\begin{equation}
\frac{dN}{dS}=\left(\frac{N_{0}}{S_{0}}\right)\left(\frac{S}{S_{0}}\right)^{-\gamma}\mathrm{e}^{-S/S_{0}}.
\end{equation}
\noindent
We adopted parameters obtained by \citet{casey2013} from a fit to the number counts in a roughly 0.1\,deg$^{2}$ portion of the COSMOS field, namely $N_{0}=3300$\,deg$^{-2}$, $S_{0}=3.7$\,mJy and $\gamma=1.4$. While \citet{geach2016} also fit this model to their number counts, we found that the above values were more consistent with our data as they predicted more bright sources. Positions were randomly selected to simulate Poisson statistics, with no clustering. The maps were convolved with a nominal SCUBA-2 beam, and Gaussian noise was added in order to produce the equivalent rms in each fields' actual map. The maps were then smoothed again with the SCUBA-2 beam,  which is the matched filter that optimizes point-source detection \citep[see][Appendix A]{chapin2011}.

SMA follow-ups were simulated by locating all peaks in the map brighter than a certain cutoff, which was determined to be the faintest SCUBA-2 source targeted by our actual SMA observations in a given field. The mock SMA follow-ups were performed by creating 9\,arcsec $\times$ 9\,arcsec thumbnail images centred on a bright SCUBA-2 source's peak pixel; we chose 9\,arcsec as a characteristic thumbnail size, since beyond this radius we no longer expect to be seeing the source/sources that contribute to the SCUBA-2 flux density we are following up. The thumbnail images were smoothed by a 2.4-arcsec FWHM beam, which accurately reconstructed our actual SMA observations because most of the galaxies in our data are unresolved. The distribution of pixel flux densities from all of the mock SMA observations is then a good estimator for the prior, since it takes into account both resolution and selection effects present in our observations. For each of the five fields where we have data, we repeated our simulation a sufficient number of times to obtain good statistics. The parameters used in each of the five fields' simulations are summarized in Table \ref{table-sim}.

\begin{figure}
\includegraphics[width=\columnwidth]{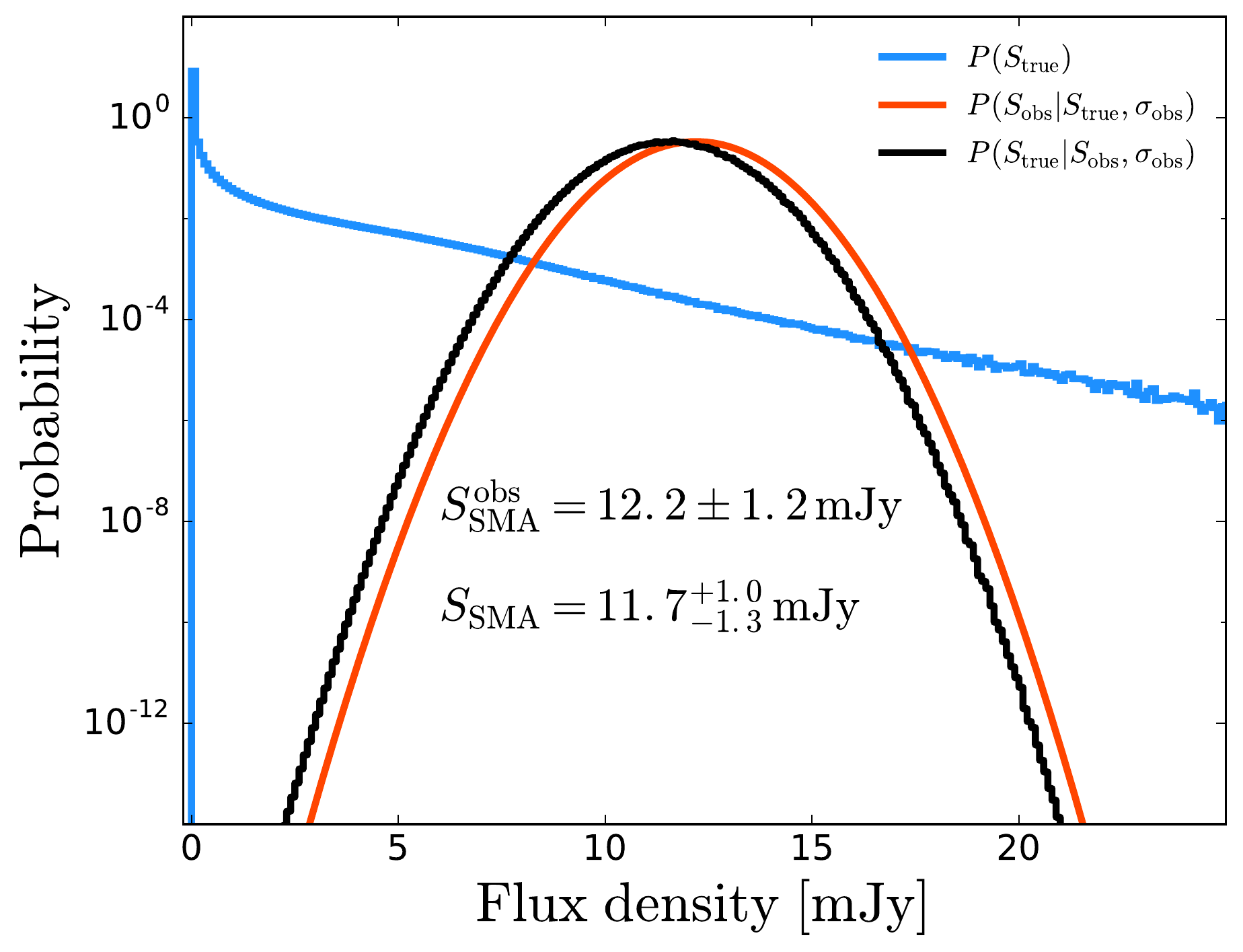}
\caption{Probability distributions for the flux density of COSMOS14, a typical source in our data set. The blue curve is the prior, which is calculated by binning pixels resulting from simulating S2CLS fields and making small SMA thumbnail images centred on the brightest sources. The red curve shows the flux density of COSMOS14 measured from our data, where the uncertainty is assumed to be Gaussian. The black curve is the posterior probability distribution, which peaks at a slightly lower, deboosted flux density value due to the presence of many more faint galaxies in the simulated sky. The deboosted flux density uncertainties given represent a 68 per cent confidence interval about the peak.}
\label{deboost}
\end{figure}

We constructed a posterior probability distribution for the intrinsic flux density of each source using priors from their respective fields. In Tables \ref{table1}--\ref{table5} under the column $S_{{\rm SMA}}$ we report the deboosted flux density as the peak in the posterior probability distribution, and we give error bars representing 68 per cent confidence intervals. In Fig.~\ref{deboost} we show an example of this deboosting technique for a typical source, COSMOS14, which, according to our simulations, is expected to be 4 per cent fainter than indicated by our maps. Note that the error bars do not necessarily increase, but the signal always decreases so that the S/N always decreases. We also note that COSMOS22, which had a S/N value just at the threshold of 4.0, had a probability density function that also peaked at zero flux density, so we report a 68 per cent upper limit for this source as well.

Cases where a single bright SCUBA-2 source is resolved into two or more faint galaxies are more difficult to deboost. In our simulations we do not include any galaxy-galaxy interactions, clustering or lensing, and we only follow-up the SCUBA-2 sources brighter than a certain threshold, so we cannot use our approach to obtain deboosting fractions for those faint galaxies which contribute to single, bright SCUBA-2 peaks. For example, should a bright SCUBA-2 source resolve into one bright SMG above our follow up threshold and one or more faint SMGs below our follow up threshold, our boosting correction would be applicable only to the bright SMG. We therefore define all faint galaxies to be those with flux densities 1\,mJy less than the cutoff used to determine which SCUBA-2 sources were to be followed up by the SMA in our simulations in a given field. Galaxies LHN13a and LHN13b resolved completely from a SCUBA-2 peak and are considered faint, while COSMOS11b resolved from a SCUBA-2 peak along with a bright companion. We did not correct the measured flux densities for these SMGs, and we simply use the measured values throughout the paper; in the $S_{{\rm SMA}}$ column of Tables \ref{table1}--\ref{table5}, we report a value of N/A for these cases. We note that neglecting to deboost these faint sources will have no effect on the bright end of the number counts.

\begin{table}
\caption{Parameters describing our simulations, which we use to calculate the expected level of flux boosting in our measurements.}
\begin{tabular}{lcccc|}
\hline                                 
Field & S2CLS area & S2CLS noise & SMA cutoff  \\
 {} & [deg$^{2}$] & [mJy] & [mJy] \\
 \hline
 UDS & 0.96 & 0.9 & 7.8 \\ 
SSA22 & 0.28 & 1.2 & 6.7 \\ 
COSMOS & 2.22 & 1.6 & 7.2 \\ 
LHN & 0.28 & 1.1 & 8.1 \\ 
EGS & 0.32 & 1.2 & 9.8 \\ 
 \hline
\end{tabular}
\label{table-sim}
\end{table}

\subsection{Comparison with the parent SCUBA-2 sample}\label{comparison}

The accuracy with which SCUBA-2 sources can be localized is well understood to be a function of observed S/N (assuming no multiplicity), and is approximated as \citep[equation B22 in][]{ivison2007}

\begin{equation}
\label{separation}
\Delta\alpha=\Delta\delta=0.6 \ {\rm FWHM} \ [{\rm (S/N)_{obs}^2}-(2\beta+4)]^{-1/2},
\end{equation}
\noindent
where FWHM is the beamsize of SCUBA-2 and $\beta$ is the local slope of the cumulative number count used as a prior to correct the observed flux densities for boosting. To examine the positional accuracy of our sample we computed the radial distance between our interferometrically-detected sources and those of the parent SCUBA-2 catalogue as a function of the detected S/N from \citet{geach2016}. For cases where multiple SMA/ALMA sources are detected, we simulated a simple (noiseless) SCUBA-2 image by convolving point sources at the SMA positions with a nominal SCUBA-2 beam with a FWHM of 14.8\,arcsec and calculated the location of the peak intensity, which is then compared to the reported SCUBA-2 source position. We took into account offsets between the SMA and SCUBA-2 reference frames on a field-by-field basis by subtracting the mean difference in right ascension and declination from each calculated offset. 

\begin{figure}
\includegraphics[width=\columnwidth]{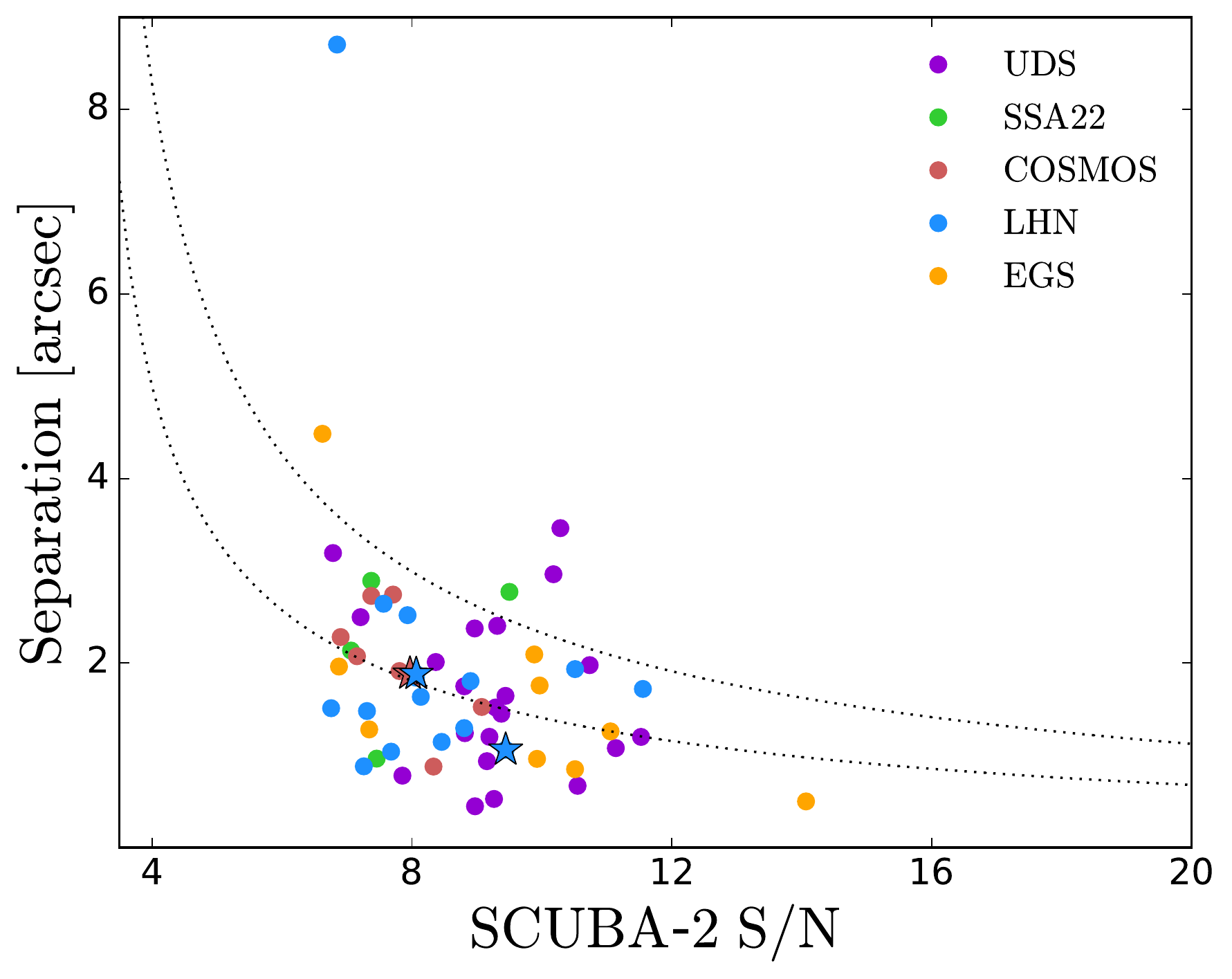}
\caption{Radial offset of SMA-detected sources from their SCUBA-2 counterparts. Where multiple counterparts are detected we smooth the sources with the nominal SCUBA-2 beam and locate the peak flux density and compare this to the given SCUBA-2 position. These sources are highlighted in the figure by stars. Also shown are the expected 68 per cent and 95 per cent positional uncertainties as a function of detected S/N for SCUBA-2.}
\label{position}
\end{figure}

In Fig.~\ref{position} we plot the radial separation of our SMA positions relative to the SCUBA-2 positions (except for the 13 sources where we did not detect a galaxy) as a function of SCUBA-2 S/N. Also shown are theoretical 68 per cent and 95 per cent contours, derived using Equation \ref{separation} with $\beta=2.4$ (note that 68 per cent and 95 per cent contours are actually at $1.51\sigma$ and $2.50\sigma$, respectively, since the probability density is proportional to $r \mathrm{e}^{-r^{2}/2\sigma^{2}}$). Five sources lie above the 95 per cent contour, corresponding to about 7 per cent of the sources in our sample, which is only marginally more than expected. The typically small positional offsets imply that in most of our maps the SMA primary beam corrections are not very significant. There also appears to be one outlier with a 9\,arcsec offset, LHN09, which could be a misidentification.

\begin{figure}
\includegraphics[width=\columnwidth]{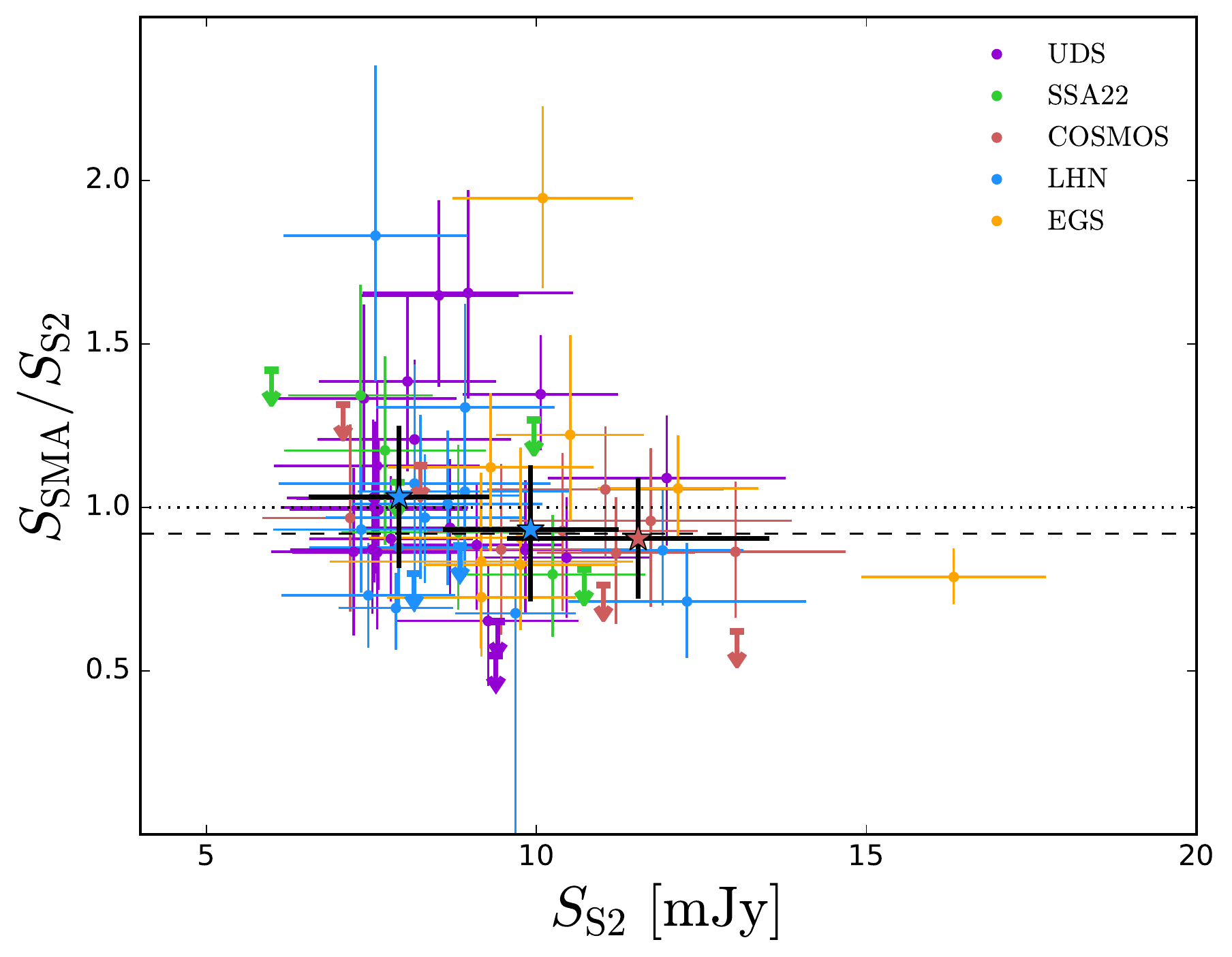}
\caption{Comparison of the SCUBA-2 deboosted flux density from \citet{geach2016} to the ratio of our SMA deboosted flux densities to each corresponding SCUBA-2 flux density. Where a single CLS source is resolved into multiple components, we have summed each components' flux density weighted by the SCUBA-2 beam response. These sources are shown as stars. Targets were we were only able to determine upper limits on the SMA flux density are shown as downward pointing arrows. The dotted line indicates a flux ratio of 1, expected if we recovered 100 per cent of the SCUBA-2 flux density, and the dashed line shows our median ratio of $0.93^{\scriptscriptstyle +0.05}_{\scriptscriptstyle -0.03}$, which could be less than 1 due to the presence of faint galaxies below the sensitivity of our observations.}
\label{flux}
\end{figure}

Next we compare the interferometric flux density observations to those from SCUBA-2 to check the reliability of the flux densities in our data set. We use the boosting-corrected flux densities reported by \citet{geach2016} and our boosting-corrected flux densities. When comparing the cases where a SCUBA-2 source is resolved into multiple components in our SMA data, we again simulated a simple SCUBA-2 image by convolving point sources at the SMA positions with a 14.8\,arcsec beam, then evaluate the flux density at the reported SCUBA-2 position. The results are shown in Fig.~\ref{flux}, where we have plotted $S_{{\rm S2}}$ versus $S_{{\rm SMA}}$/$S_{{\rm S2}}$. In this figure the multiples are shown as stars and the 13 sources where we were only able to obtain upper flux density limits in our SMA observations are shown as downward pointing arrows.

The median value of the ratio $S_{{\rm SMA}}$/$S_{{\rm S2}}$, not including the 12 blank maps, is $0.95^{\scriptscriptstyle +0.05}_{\scriptscriptstyle -0.04}$, where the uncertainty was calculated as the 68 per cent confidence interval from bootstrap resampling. When we add the ratios of $S_{{\rm SMA}}$/$S_{{\rm S2}}$ from the blank maps using the upper limits derived for $S_{{\rm SMA}}$, the median value of $S_{{\rm SMA}}$/$S_{{\rm S2}}$ is $0.95^{\scriptscriptstyle +0.04}_{\scriptscriptstyle -0.04}$, which is almost identical to the previous estimate. Considering the additional calibration uncertainty of order 10 per cent between the two instruments, we cannot place any useful limits on missing flux densities from faint sources as our SMA data is not deep enough.

\subsection{Completeness}\label{completeness}

Here we discuss in detail the completeness of our observations with respect to the parent S2CLS survey. Since our targets were selected based on early S2CLS maps with higher noise, it is important to understand our sample in terms of the final, published maps. In addition, we have to decide how many sources from other experiments we wish to include in our sample, since they will affect our completeness. 

The latter question is important because we have not specifically targeted any sources below the given flux density limit (which depends on the field); however, images from \citet{younger2007,younger2009}, \citet{aravena2010}, \citet{tamura2010}, \citet{smolcic2012b,smolcic2012} and \citet{simpson2015} extend much deeper, and in several cases a faint source that would have been omitted from our study turned out to be bright enough to affect the bright end of the number counts when observed by the SMA or ALMA. Including sources like this could potentially bias our results, since our survey would then not really be `blind', rendering the analysis much more difficult to interpret. 

Our approach to this problem involves two steps. First, we incorporate into our catalogue all sources from these works that have SCUBA-2 deboosted flux densities greater than the faintest source we targeted in our observations in a given field (see Table \ref{table-sim}, under the `SMA cutoff' column). These sources are included in Tables \ref{table1}--\ref{table5}, and we have used the naming conventions given in their respective papers. There are seven sources from \citet{younger2007}, four from \citet{younger2009}, one from \citet{aravena2010}, one from \citet{tamura2010}, one from \citet{smolcic2012b}, four from \citet{smolcic2012} and 16 from \citet{simpson2015}, for a total of 34 archival sources. Since the the five sources from \citet{smolcic2012b,smolcic2012} were observed at 1.3\,mm, we convert the reported flux densities to 860\,$\mu$m by modelling a typical submm galaxy spectral energy distribution as a modified blackbody function with a temperature fixed to 35\,K, a dust emissivity index of 2 and a redshift of 2. We note that the flux densities from \citet{younger2007,younger2009}, \citet{aravena2010}, \citet{tamura2010} and \citet{smolcic2012b,smolcic2012} have not been corrected for flux boosting, so we use their direct measurements; the flux densities from \citet{simpson2015} have been corrected for flux boosting, which are given under the $S_{{\rm ALMA}}$ column, and we use these values for further analysis. We also include in our work the SMA observation of Orochi from \citet{ikarashi2011}, the gravitational lens in the UDS field. This brings the total number of archival sources to 35, and the total number of samples in our analysis to 105.

Next, we calculate a completeness level for that field by dividing the total number of SCUBA-2 sources targeted in our sample by the total number of SCUBA-2 sources in the parent sample in a given flux density bin. We looked at bins above 8\,mJy with widths of $\Delta S=1$\,mJy. In this way we are effectively treating the external sources as if we had targeted them ourselves, introducing as little bias as possible, while still using all of the data. We can then use the calculated completeness values in each bin to correct for the missing sources introduced in the final, deeper S2CLS SCUBA-2 maps.

In the UDS field, we have targeted sources down to SCUBA-2-deboosted flux densities of 7.2\,mJy. After introducing the sources from \citet{simpson2015} with deboosted SCUBA-2 flux densities greater than 7.2\,mJy, we find that our catalogue reaches a completeness of 96 per cent for $S_{850}\,{>}\,8$\,mJy, where the unobserved 4 per cent of sources are cases where a SCUBA-2 flux density was scattered to a higher value with the additional exposure time. At fainter flux densities our completeness falls below 80 per cent, which we deem to be too low to be used reliable, and in the brighter regime of $S_{850}\,{>}\,9$\,mJy we have 100 per cent completeness. A similar analysis performed for the ALMA sources observed by \citet{simpson2015} resulted in completeness levels of 50 per cent for $S_{850}\,{>}\,8$\,mJy, 56 per cent for $S_{850}\,{>}\,9$\,mJy, and 73 per cent for $S_{850}\,{>}\,10$\,mJy, which shows that our observations offer a significant improvement in this field owing to the fact that our targets were selected from later versions of the S2CLS maps. 

In the SSA22 field we have followed up 100 per cent of the sources with a deboosted SCUBA-2 850-$\mu$m flux density greater than 10\,mJy. In this field there are no sources with SCUBA-2 850-$\mu$m deboosted flux densities between 9 and 10\,mJy, and below 9\,mJy our data do not cover enough sources to allow us to reliably estimate the number counts. Despite the fact that we have targeted five additional sources less than the $S_{850}\,{=}\,10$\,mJy level, two sources scattered up to about 8\,mJy in the deeper SSA22 850\,$\mu$m S2CLS map after our targets were selected, and so our completeness for $S_{850}\,{>}\,8$\,mJy is only 57 per cent.

In the COSMOS field, only about 50 per cent of the total area was mapped to a nominal depth of 1.6\,mJy in the published 850\,$\mu$m S2CLS maps used in our study, and the remaining half has recently been completed (S2COSMOS: Simpson et al.~in preparation); our completeness calculation for this field is based on the current data available in \citet{geach2016}. We find that, with the addition of the observations from \citet{younger2007,younger2009}, \citet{aravena2010} and \citet{smolcic2012b,smolcic2012} down to 7.1\,mJy, our faintest target, we have completeness of 89 per cent for $S_{850}\,{>}\,10$\,mJy, and 100 per cent completeness for $S_{850}\,{>}\,11$\,mJy. Below 10\,mJy our sample becomes very sparse. There are two sources with deboosted SCUBA-2 850\,$\mu$m flux densities of 10.0 and 10.1\,mJy that have not been observed with the SMA in our campaign, nor in any of the literature, due to their low S/N in earlier SCUBA-2 and AzTEC maps.

We have fully probed the LHN field down to $S_{850}\,{=}\,7.5$\,mJy, achieving 100 per cent completeness. Below this we targeted one source whose corresponding deboosted SCUBA-2 850$\mu$m flux density is 7.3\,mJy, but we do not try to probe number counts this low.

Lastly, our sample does not include any EGS sources with $S_{850}\,{<}\,9$\,mJy, while for $S_{850}\,{\geq}\,9$\,mJy we have resolved all of the available S2CLS sources, and thus every detection is statistically significant for estimating the counts in this field.

We now consider the completeness of our total data set. We have observed nearly all sources down to 850\,$\mu$m flux densities of 10\,mJy in these five cosmological fields, reaching a completeness level of 95 per cent for $S_{850}\,{>}\,10$\,mJy. As described above, there are two SCUBA-2 sources with deboosted flux densities at 850\,$\mu$m of 10.0 and 10.1\,mJy that have no interferometric data, both in the COSMOS field. When considering our full data set, these two sources comprise 5 per cent of the total number of sources with $S_{850}\,{\geq}\,10$\,mJy. In Table \ref{completeness_table} we summarize our completeness calculations for each field, for $S_{850}\,{>}\,8$\,mJy and $S_{850}\,{>}\,10$\,mJy.

\begin{table}
\caption{Completeness levels calculated for each field in our study, as well as for the total data set.}
\begin{tabular}{lccc|}
\hline                                 
Field & Completeness & Completeness \\
 {} & $S_{850}\,{>}\,8$\,mJy & $S_{850}\,{>}\,10$\,mJy \\
 \hline
 UDS & 96$\%$ &  100$\%$ \\ 
SSA22 & 57$\%$ &  100$\%$ \\ 
COSMOS & 56$\%$ &  89$\%$ \\ 
LHN & 100$\%$ &  100$\%$ \\ 
EGS & N/A &  100$\%$ \\ 
\hline
Total & 77$\%$ &  95$\%$ \\ 
 \hline
\end{tabular}
\label{completeness_table}
\end{table}

\section{Results and discussion}\label{discussion}

\subsection{Number counts}

We now estimate the cumulative number counts of our sample of interferometrically-detected SMGs. Our calculations are restricted to counts within the completeness regimes discussed in Section \ref{completeness}. The areas for each field are given in Table \ref{table-sim} from \citet{geach2016}, totalling 4.06\,deg$^{2}$ for our complete survey. We calculate the cumulative number count in bins of $\Delta S=1$\,mJy by simply counting the total number of sources with $S_{860}\,{>}\,S$ and dividing by the total area. To correct for incompleteness at the fainter flux density bins, we multiplied the total area by the fraction of sources targeted in our survey relative to the sources in the parent S2CLS catalogue (i.e. the completeness from Table \ref{completeness_table}).

For the 12 observations where only upper limits were obtained for the SMA counterparts we use the upper limit flux density as the deboosted SMA flux density; all $4\sigma$ upper limits we have measured constrain the flux densities of these sources to $S_{860}{<}\,10$\,mJy, below the regime where we are calculating the counts, so we are not introducing any bias in the flux density region studied in this work by doing this. The source SSA22-04 is however an exception, where we have constrained the flux density to be less than 12.6\,mJy but the corresponding SCUBA-2 flux density is 10.0\,mJy. Since our SMA observations of this source have not been able to provide any further information, we have removed this source from our calculation and corrected for the incompleteness this introduces using the procedure described above. Lastly, for plotting purposes, we remove all repeated points, that is, points where there is no change in the cumulative number count in two adjacent bins because there are no sources between $S$ and $S+\Delta S$.

\begin{figure*}
\includegraphics{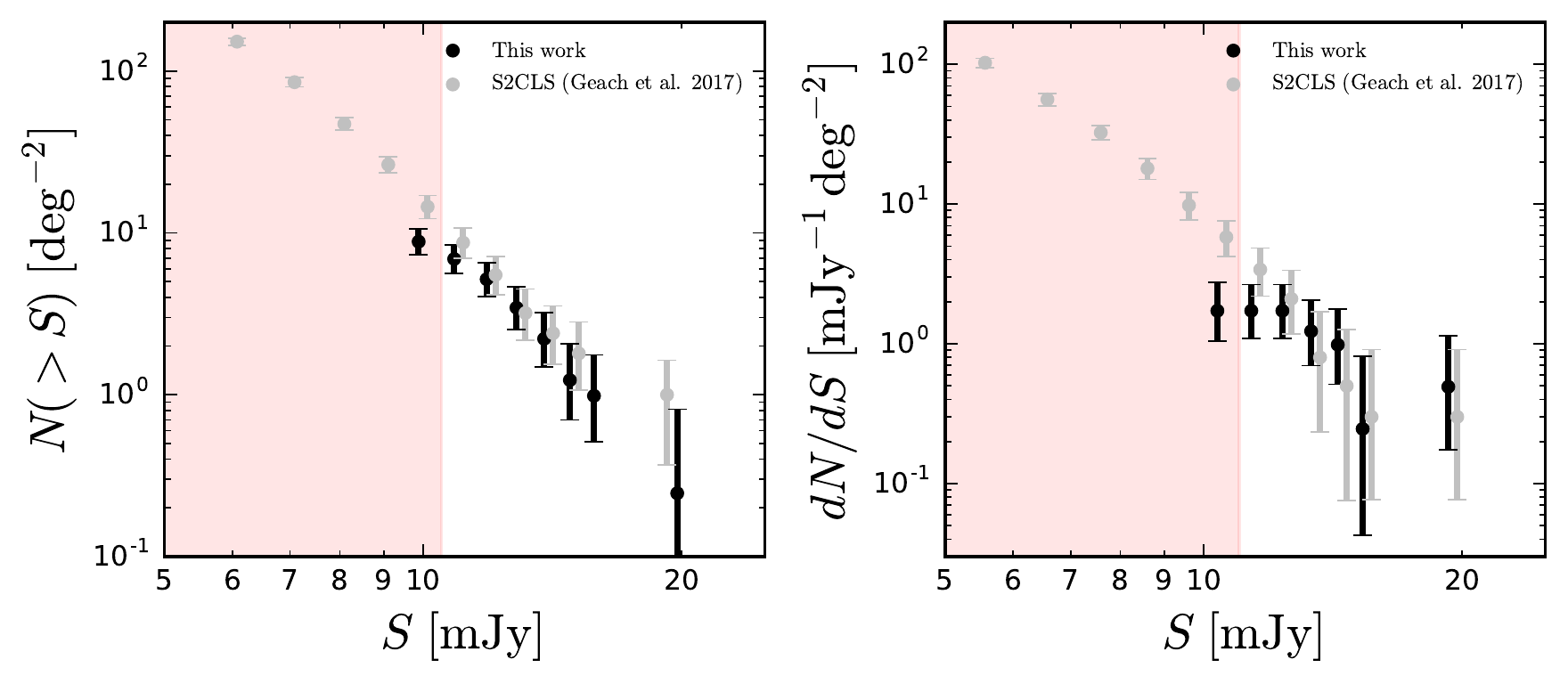}
\caption{Cumulative and differential number counts derived from our data set. The single dish results from the S2CLS \citep{geach2016} are shown for comparison. Values are slightly offset from each other in each bin for clarity. The shaded region marks where our data is no longer 100 per cent complete. An offset between our results of 20 to 30 per cent is seen in the cumulative count, although the points overlap within the uncertainties.}
\label{numcount}
\end{figure*}

The results for the cumulative number count from our full survey are shown in Fig.~\ref{numcount}. The error bars are calculated as 68 per cent confidence intervals from Poisson statistics \citep[see][]{gehrels1986}. In addition, we show the S2CLS cumulative count results from \citet{geach2016} for comparison. We have also shaded the boundary marking the 100 per cent completeness of our sample (set by the COSMOS field).

We then compute the differential number counts, following the same procedure as above. The results are also shown in Fig.~\ref{numcount}, together with the S2CLS differential counts from \citet{geach2016} and the region marking the boundary of 100 per cent completeness.

We also calculated cumulative and differential number counts separately for each field. Overall we saw no significant field-to-field variations, although the counts in the smaller field were quite uncertain due to Poisson noise. We concluded that there was no additional information to gain from a field-by-field analysis and therefore only discuss the counts from the full survey in the following sections.

\begin{figure*}
\includegraphics{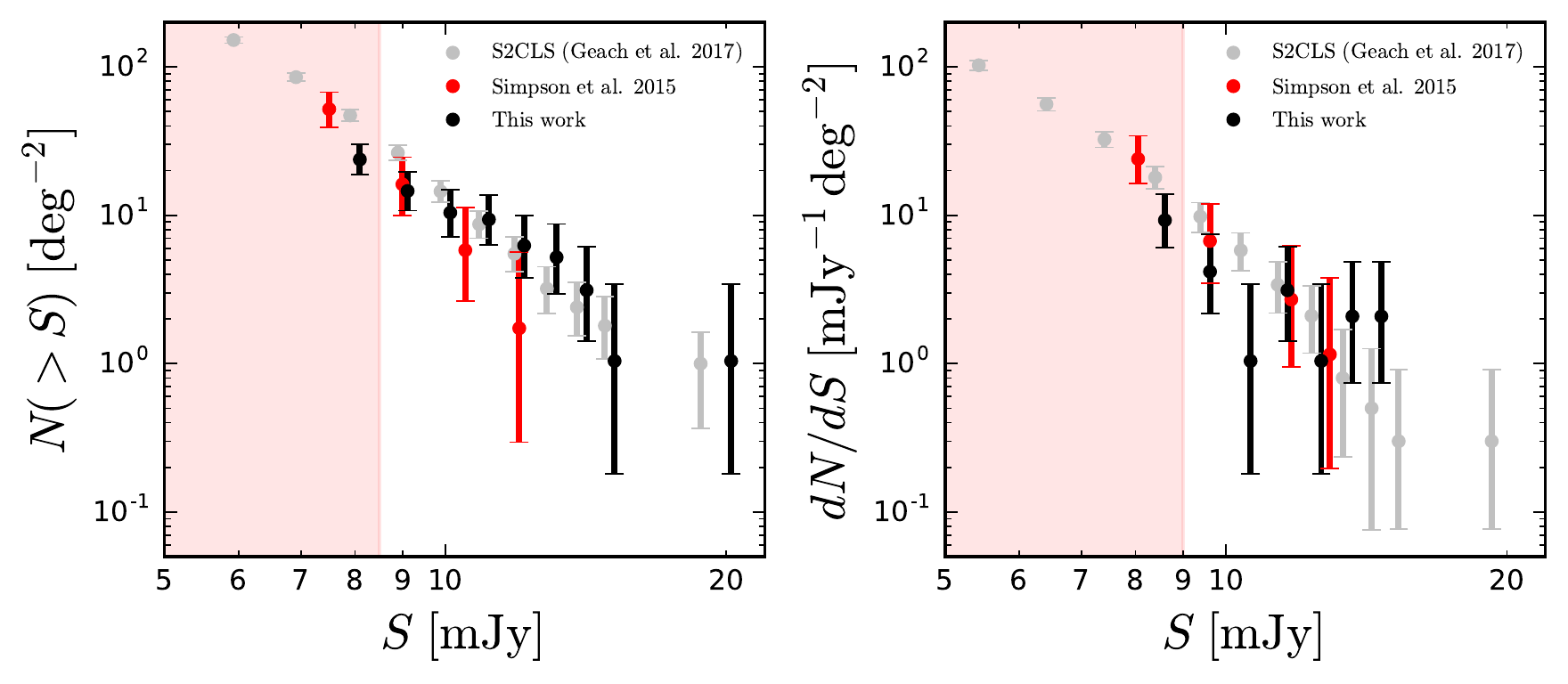}
\caption{Cumulative and differential number count comparison for the UDS field. The results from \citet{simpson2015}, derived from a smaller sample of the full parent S2CLS catalogue of the UDS field, are shown in red, alongside our more complete sample in black, where we have used only data from the UDS field as well. The results broadly agree, although we see evidence for less bright sources in the \citet{simpson2015} sample. Also shown as the shaded region is where our data is not 100 per cent complete; our UDS data is 96 per cent complete for $S{>}\,8$\,mJy.}
\label{uds}
\end{figure*}

In Fig.~\ref{uds} we show our cumulative and differential number counts for the UDS field alone compared to those derived by \citet{simpson2015}, along with the shaded region indicating our 100 per cent completeness limit. There seems to be a slight lack of sources at $S_{850}\,{\gtrsim}\,10$\,mJy seen by \citet{simpson2015}, however there are three SCUBA-2 sources (UDS03, UDS08 and UDS09) that were not targeted in their work as they did not appear to among the brightest 30 UDS sources in the earlier, shallower S2CLS maps used to design their follow-up ALMA programme. Also shown in Fig.~\ref{uds} is the cumulative and differential count from the S2CLS data in \citet{geach2016}. By including the three bright UDS sources to the number counts we find no strong evidence for diagreement between the single-dish measurements from \citet{geach2016}, the measurements from \citet{simpson2015} and our work within the uncertainties and within the overlapping flux density regimes.

Similar single dish counts were also obtained by the LESS survey \citep{weiss2009}, which was a 0.35-deg$^{2}$ 870-$\mu$m survey of the E-CDF-S carried out with LABOCA, which has a FWHM of 19.2\,arcsec. The LESS survey detected a total of 126 submm galaxies to a noise level of approximately 1.2\,mJy. Following this, a high-resolution follow-up campaign was carried out by \citet{hodge2013} using ALMA, and the number counts were presented by \citet{karim2013}. They found no sources brighter than $S_{870}\,{\simeq}\,9$\,mJy despite there being 12 LABOCA sources in this regime, implying a cut-off to possible FIR luminosities and star-formation rates. 

\begin{figure*}
\includegraphics{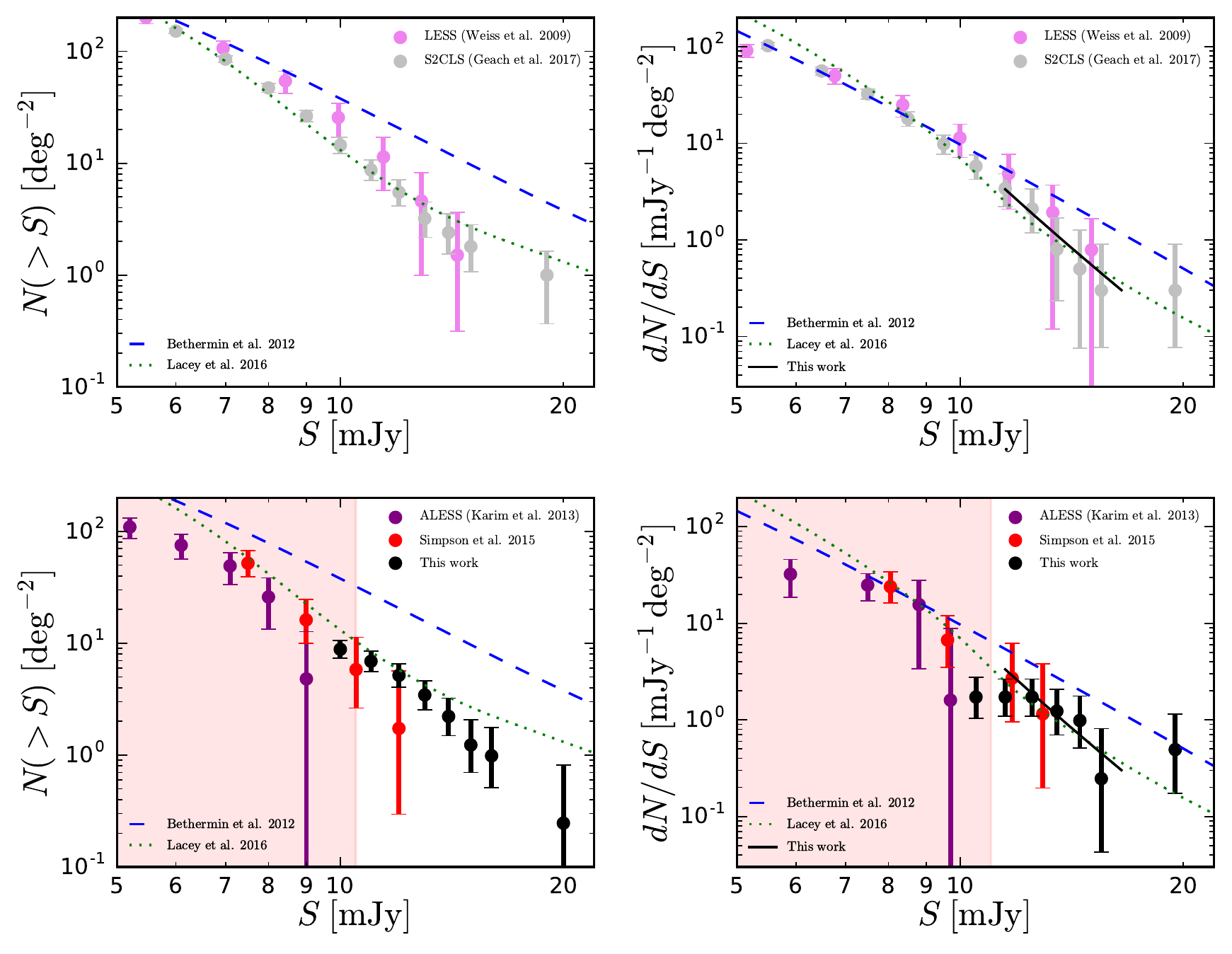}
\caption{Cumulative and differential number counts for the two large single dish submm surveys LESS \citep{weiss2009} and S2CLS \citep{geach2016} on the top row.  On the bottom row we show cumulative and differential number counts from \citet{karim2013} and \citet{simpson2015}, interferometric follow-up studies of the LESS and S2CLS surveys, respectfully, along with our SMA results and the shaded region indicating where our data is no longer 100 per cent complete. Also shown are the models of \citet{bethermin2012} and \citet{lacey2016}. The black solid line shows the best-fit power to our differential distribution between 11 and 16\,mJy.}
\label{final_comparison}
\end{figure*}

We compare our results to these earlier works in Fig.~\ref{final_comparison}, where on the top row we have plotted the cumulative and differential number counts from LESS and the S2CLS (i.e. two single dish submm surveys), and on the bottom row we have plotted the cumulative and differential number counts from \citet{karim2013}, \citet{simpson2015} and our work (i.e. high angular resolution follow-up studies); the shaded region indicating where our data is no longer 100 per cent complete is shown as well. In this plot we have included the number counts from models of evolving star-forming galaxies, specifically the empirical model from \citet{bethermin2012} and the GALFORM model from \citet{lacey2016}.

We see no evidence for a lack of high flux density sources, as hinted at by the results of \citet{karim2013}, and instead see the number count carrying on without a steep drop-off to around 15\,mJy, in agreement with the counts from \citet{simpson2015}. On the other hand, our cumulative number count is systematically lower than the parent S2CLS cumulative number count, which can be readily seen in Fig.~\ref{numcount}. We calculate the mean fractional difference between the two cumulative number counts between 11 and 15\,mJy to be $14\,{\pm}\,6$ per cent, where the uncertainty is the standard error of the mean. The origin of this difference can likely be attributed to the blank maps in our data sample, which overall reduce the total number of bright sources used to calculate the cumulative number count. This is likely a consequence of multiplicity in the SCUBA-2 sample, and will be discussed in more detail in the following section.

Lastly, we fit a power law to our differential count in order to quantitatively determine the steepness of the counts in the high flux density regime. We fit only points between 11 and 16\,mJy, since our flux density coverage for smaller values is not 100 per cent complete, and beyond 16\,mJy the differential number count begins to flatten and are not well-described by a simple power law. Our model is of the form

\begin{equation}
\frac{dN}{dS}=N_{0} S^{-\gamma},
\end{equation}
\noindent
and we find best-fit parameters of $\gamma=5.3\pm1.8$ and $N_{0}=(0.9\pm4.1)\times10^{6}$\,mJy$^{-1}$\,deg$^{-1}$. This best-fit curve is plotted alongside our data in Fig.~\ref{final_comparison}. 


\subsection{Multiplicity}

The importance of galaxy interactions and mergers for the intense star-formation rates observed in many submm galaxies is a hotly debated topic. Here we discuss the multiplicity seen in our large sample of bright, 850-$\mu$m-selected galaxies at a resolution of about 2\,arcsec, and contrast our observations with previous works. 

There is first the question of how to precisely define a multiple; for example, \citet{lambas2012} defined multiples by their flux density ratio, and pairs with brightest to second-brightest ratios less than 3 were considered multiples, since this value provides a reasonable cut-off for finding single dish sources whose flux densities have been seriously affected. Our observations are not able to detect ratios as high as 3 but we have probed the regime of ratios close to 1, where single dish flux densities are the most seriously affected. This means that our observations will miss all but the closest pairs in terms of flux density ratios. Thus our observations are not sensitive enough to measure the intrinsic multiplicities of the brightest submm sources, but instead determine the fraction of the brightest submm sources that are fainter by a factor of about 2.

There is also the question of how to interpret our 12 blank maps. This may in several cases be attributed to faint multiples being missed due to the noise level in our SMA observations; for example, using more sensitive ALMA observations, \citet{simpson2015} found two cases of bright, 7--11\,mJy SCUBA-2 850\,$\mu$m sources resolving into multiple $<6$\,mJy sources at 870\,$\mu$m, which would not be detected in most of our SMA pointings. It is thus plausible to attribute our null-detections to cases where the SCUBA-2 blended source is composed of multiple faint sources that are lost in the noise. However, this interpretation must be treated as an upper limit to the multiplicity number as some blank maps in our data may just be SCUBA-2 sources that are fainter than reported in \citet{geach2016} due to flux calibration uncertainties between the two measuremenrs and not multiples. We also have to deal with the fact that there also are instances where the 4$\sigma$ flux density threshold in our SMA maps is greater than the flux density of the SCUBA-2 source we are trying to detect and so we wouldn't expect to see even a single bright source. In these cases we cannot claim evidence for detecting multiplicity. Specifically, UDS14, UDS15, SSA22-03, COSMOS06, COSMOS17, LHN11 and LHN12 each have SMA flux density limits less than their observed SCUBA-2 counterpart flux densities, and so may be composed of multiple galaxies below our $4\sigma$ limit, whereas for SSA22-04, SSA22-07, SSA22-09, COSMOS22 and COSMOS25 we are not able to say anything about the galaxies contributing to the SCUBA-2 flux density.

Lastly, in order to properly incorporate the interferometric observations taken from the literature into our study we must only count multiples consistent with our definition of the multiplicity detectable with our SMA observations, namely groups that have flux density ratios close to 1. Since we also interpret blank maps as multiples, we will also consider observations that found several galaxies all fainter than 6\,mJy, the average depth of our data, as sources that would be seen as multiples. UDS156.0/UDS156.1, UDS57.0/UDS57.1 and COSLA-23-N/COSLA-23-S satisfy the former criteria, while UDS286.0/UDS286.1/UDS286.2/ UDS286.3 and UDS199.0/UDS199.1 satisfy the latter criteria. Note that AzTEC11.N/AzTEC11.S is not considered a multiple under this criteria. 

We can now calculate an upper limit to the fraction of bright submm sources that resolve into two or more galaxies with flux ratios close to 1, effectively decreasing the single-dish submm source flux density estimate by a factor of around 2. We have observed three multiples in our SMA data and inferred a further seven multiples from blank maps, and taken five more multiples from the literature. Out of 105 observations (removing the gravitational lens Orochi), this results in a fraction of ${\lesssim}\,0.15$.

The question of wheather the multiplicity seen in our SMA images correspond directly to galaxy mergers is difficult to address with our data. First, we note that the physical scale being probed by the SMA's resolution, namely 2.4\,arcsec, at a fiducial redshift of $z\,{=}\,2$ is about 20\,kpc, which is around the same separation seen with major-mergers in the local Universe \citep[e.g.][who examined a set of about 2000 galaxy pairs at $z\,{<}0.1$]{lambas2012}. On the other hand, with enough sensitivity one will almost always detect faint multiples; it has been suggested that line-of-sight projections could account for a significant fraction of the multiplicity seen in bright SMGs \citep{cowley2015}. However, under the assumption that a given multiple in our SMA data are physically associated, and given the observation that the ultra-luminous galaxies probed in our survey are rare, it seems likely that detecting two or more of these rare galaxies would be much more probable in significantly denser regions of the early Universe, such as in proto-clusters, than compared to the field. The multiples detected in our survey, being composed of groups of equally bright galaxies, could therefore be useful markers of massive galaxy proto-clusters around redshifts of 2.

\subsection{Density of extremely luminous galaxies}

Our sample of galaxies represent some of the most luminous and intensely star-forming sites in the Universe. In order to estimate the average luminosity and SFR of our sample of galaxies, we take the set of 52 (rest frame) SEDs provided by \citet{danielson2017} from galaxies in the ALESS and calculate their mean to derive an average SMG SED. We then fix the redshift at $z\,{=}\,2$ and set the normalization to get 10\,mJy at 860\,$\mu$m, and integrate this model from 8 to 1000\,$\mu$m (which is the definition of the IR luminosity). This calculation results in $4.5\,{\times}\,10^{12}$\,L$_{\sun}$, and this can be converted to a SFR using the relationship from \citet{kennicutt98} modified for a Chabrier IMF \citep{chabrier03} (i.e. SFR[M$_\odot$\,yr$^{-1}$]\,=\,9.5$\times$10$^{-11}$\,L$_{\rm{IR}}/$L$_\odot$) to yield ${\sim}\,400$\,M$_{\sun}$\,yr$^{-1}$. 

Using the above result we can recast our number counts in terms of intrinsic SFRs. Since we see a surface density of galaxies brighter than 10\,mJy of $8^{+2}_{-1}$\,deg$^{-2}$, this is a good approximation for the number of galaxies with SFRs ${\gtrsim}\,400$\,M$_{\odot}$\,yr$^{-1}$. Here we are assuming that none of our sources are being gravitationally lensed, which would reduce their intrinsic SFRs; we will address the fraction of gravitaionaly lensed galaxies in our sample in future work. Assuming half of our sources lie between $z\,{=}\,2$--3 and using the cosmological parameters from \citet{planck2014-a15}, this implies a likely volume density of ${\sim}\,3^{+0.7}_{-0.6}\,{\times}\,10^{-7}$\,Mpc$^{-3}$. 

The lifetimes of starbursts in SMGs are expected to be of order 100\,Myr  \citep[e.g.,][]{swinbank2006,hainline2011,hickox2012,bothwell2013}. Between $z\,{=}\,3$ and $z\,{=}\,2$ the lookback time is approximately 1\,Gyr, which implies that the volume density galaxies descended from this population is larger by a factor of 10, or ${\sim}\,3\,{\times}\,10^{-6}$\,Mpc$^{-3}$. This number density can be compared to the number density of local ($z\,{<}\,0.1$), red quiescent galaxies, which are the expected descendants. The local volume density as a function of stellar mass (i.e. the stellar mass function) of quiescent galaxies has been measured by selecting `red sequence' galaxies based on their colour \citep{bell2003,baldry2012} and by selecting `star-forming sequence' galaxies based on their location on a SFR-stellar mass diagram \citep{moustakas2013}, and both techniques agree well for stellar masses ${\gtrsim}\,10^{11}$\,M$_{\sun}$. We have taken the stellar mass function of \citet{moustakas2013}, which probed the largest stellar masses out to $10^{12}$\,M$_{\sun}$ and is therefore a good comparison to our study of the most extreme galaxies, and have calculated the cumulative volume density as a function of stellar mass by integrating the stellar mass function. From this we find that local quiescent galaxies with stellar masses ${\gtrsim}\,4\,{\times}\,10^{11}$\,M$_{\sun}$ have the same volume density as the remnants of ${\gtrsim}\,400$\,M$_{\odot}$\,yr$^{-1}$ galaxies in our sample \citep[see also][]{simpson2014}. We note that the above calculation is not significantly affected by any of the assumptions we made as the stellar mass function is very steep above $10^{11}$\,M$_{\sun}$, changing by only about 0.5\,dex over 2 orders of magnitude in volume density. Adopting a fiducial values of 500\,M$_{\sun}$\,yr$^{-1}$ for the typical SFR in our sample and assuming the bursts are constant over the 100\,Myr period, this implies a stellar mass of $5\,{\times}\,10^{10}$\,M$_{\sun}$ was created during the bursts, a fraction of approximately 10 per cent the total stellar mass assembled by $z\,{=}\,0$.

\section{Summary and conclusions}\label{conclusion}

Using the SMA we have followed-up 70 of the brightest S2CLS sources spread across 4\,deg$^{2}$ in five fields. We have also included in our analysis 35 archival SMA and ALMA observations of similar nature to bring our total sample size to 105 single-dish submm sources. The synthesized beam of our observations was on average 2.4\,arcsec FWHM and the noise 1.5\,mJy\,beam$^{-1}$ as calculated from the primary beam-corrected images, sufficient to resolve the dominant SMGs contributing to the flux density peaks seen by the SCUBA-2 instrument. Altogether, we detected 62 SMGs above $4\sigma$, and saw three examples of a single SCUBA-2 peak breaking up into two or more bright SMGs. We also found that 12 of our pointings did not detect any SMGs, which may result from a SCUBA-2 peak breaking up into two or more SMGs fainter than our $4\sigma$ detection limit, which is on average 6\,mJy. 

We simulated SCUBA-2 maps and SMA follow-up pointings using the same selection criteria as for our observations in order to estimate and correct for flux boosting in our measurements. Upon applying these corrections, we found that the posterior probability distributions of two sources peaked at 0\,mJy, so we can only constrain 68 per cent upper limits on their flux densities. We tested our positional accuracy by calculating the radial distance from the peak flux density positions in our SMA images to those in the S2CLS maps, finding the spread to be consistent with the expected spread given the S/N values. We also compared our deboosted flux density measurements to the deboosted flux density measurements published in the S2CLS, and found the median ratio to be $S_{{\rm SMA}}/S_{{\rm S2}}\,{=}\,0.95 \pm 0.04$.

Assessing completeness, our sample consists of 95 per cent of the sources with $S_{850}\,{>}\,10$\,mJy with respect to the reference fields in the S2CLS, and we calculate the number counts for this regime. We compare our number counts to what was found in our parent sample, finding general agreement; however, our cumulative number count is systematically lower than the parent SCUBA-2 cumulative count by $14\,{\pm}\,6$ per cent between 11 and 15\,mJy. We also compare our counts to those from \citet{simpson2015}, who followed-up most of the bright sources in the UDS field of the S2CLS with ALMA, and find that the two estimations are in agreement. 

While multiplicity is evidently not uncommon in most of the bright single-dish sources, the effects appear not to severely affect the bright end of the number counts. We estimate an upper limit of 15 per cent for the fraction of single dish submm sources brighter than approximately 10\,mJy that resolve into two or more galaxies with similar flux densities. Instead, the most common situation involves bright single dish submm sources resolving into one slightly less bright SMG and one or more fainter ones, which only slightly lowers previous estimates of the number of bright SMGs. 

Lastly we calculate the surface density of galaxies with SFRs greater than approximately $400$\,M$_{\sun}$\,yr$^{-1}$ to be $8^{+2}_{-1}$\,deg$^{-2}$. Assuming half of the redshifts are between $z\,{=}\,2$--3, this corresponds to a volume density of ${\sim}\,3^{+0.7}_{-0.6}\,{\times}\,10^{-7}$\,Mpc$^{-3}$. Taking the typical lifetimes for starbursts to be of order 100\,Myr and noting that $z\,{=}\,2$--3 corresponds to a lookback time of about 1\,Gyr, we find a volume density of remnants to be ${\sim}\,3\,{\times}\,10^{-6}$\,Mpc$^{-3}$, which corresponds to the local volume density of quiescent galaxies with stellar masses ${\gtrsim}\,4\,{\times}\,10^{11}$\,M$_{\sun}$. Since local quiescent galaxies are expected to be descendants of the starbursting galaxies in our sample, we estimate that about 10 per cent of their total stellar mass assembled by $z\,{=}\,0$ was formed during these short bursts of star-formation.

Our observations provide the largest catalogue of the brightest interferometrically identified submm sources, probing the range of flux densities greater than 10\,mJy. This catalogue is well suited for future multiwavelength follow-up studies of some of the most extreme galaxies in the Universe. Our work with the SMA also provides some of the best available submm interferometric data of several northern cosmological fields, providing accessibility to facilities in both hemispheres.

\section*{Acknowledgements}

The James Clerk Maxwell Telescope is now operated by the East Asian Observatory on behalf of The National Astronomical Observatory of Japan, Academia Sinica Institute of Astronomy and Astrophysics, the Korea Astronomy and Space Science Institute, the National Astronomical Observatories of China and the Chinese Academy of Sciences (Grant No. XDB09000000), with additional funding support from the Science and Technology Facilities Council of the United Kingdom and participating universities in the United Kingdom and Canada. The Submillimeter Array is a joint project between the Smithsonian Astrophysical Observatory and the Academia Sinica Institute of Astronomy and Astrophysics and is funded by the Smithsonian Institution and the Academia Sinica. The authors wish to recognize and acknowledge the very significant cultural role and reverence that the summit of Maunakea has always had within the indigenous Hawaiian community. We are most fortunate to have the opportunity to conduct observations from this mountain. This work was supported by the Natural Sciences and Research Council of Canada. Ian Smail and A.~M.~Swinbank acknowledge support from the Science and Technology Facilities Council (ST/P000541/1). Ian Smail also acknowledges the European Research Council Advanced Investigator programme DUSTYGAL 321334 and a Royal Society/Wolfson Merit Award. Micha{\l} J.~Micha{\l}owski acknowledges the support of the National Science Centre, Poland through the POLONEZ grant 2015/19/P/ST9/04010. This project has received funding from the European Union's Horizon 2020 research and innovation programme under the Marie Sk{\l}odowska-Curie grant agreement No. 665778.

\bibliographystyle{mnras}
\bibliography{CLS_SMA_bib}

\appendix

\section{Data tables}

Here we provide data tables detailing our interferometric sample. Each of the five fields used in our study are summarized in a single table and ordered by decreasing deboosted SCUBA-2 flux density. The columns give the source name, the SCUBA-2 position, the SMA (or ALMA) position, the SCUBA-2 observed flux density, the deboosted SCUBA-2 flux density, the SMA (or ALMA) observed flux density and the SMA deboosted flux density. For SMA pointings that did not detect any galaxies above $4\sigma$ we provide flux density upper limits. For sources that were deboosted to 0\,mJy, we also provide $4\sigma$ upper limits. All sources are sorted by their deboosted SCUBA-2 flux density. We have used ALMA data from \citet{simpson2015} for some of the sources in the UDS field; these sources are marked with a $^{\mathrm{b}}$. We have also used SMA data from \citet{younger2007} and \citet{younger2009} for some of the sources in the COSMOS field; these sources are marked with a $^{\mathrm{c}}$ and a $^{\mathrm{d}}$, respectively. 

\begin{table*}
\caption{SMA sample plus archival ALMA data for the UDS field, ordered by decreasing deboosted SCUBA-2 flux density. The source observed by the SMA in \citet{ikarashi2011} is in bold and indicated by a $^{\mathrm{a}}$, the sources observed by ALMA in \citet{simpson2015} are in bold and indicated by a $^{\mathrm{b}}$, and all other sources were observed by the SMA in this work.}
\footnotesize
\label{table1}
\begin{threeparttable}
\begin{tabular}{lccccccc|}
\hline                                 
Source & RA/Dec SCUBA-2 & RA/Dec SMA (ALMA) & $S^{\rm obs}_{\rm S2}$ [mJy]& $S_{\rm S2}$ [mJy] & $S^{\rm obs}_{\rm SMA}$ [mJy] &  $S_{\rm SMA}$ [mJy] \\
{} & (J2000) & (J2000) & {} & {} & \kern-2em($S^{\rm obs}_{\rm ALMA}$) & \kern-2em($S_{\rm ALMA}$) \\
 \hline
 \textBF{Orochi}$^{\mathrm{a}}$ & \textBF{02:18:30.77 $-$05:31:30.8} &  \textBF{02:18:30.68 $-$05:31:31.7} & \textBF{52.7 $\pm$ 0.9} & \textBF{52.7 $\pm$ 1.2} & \textBF{\phantom{0}90.7 $\pm$ 20.7} & {}\\ 
\textBF{UDS156.0}$^{\mathrm{b}}$ & \textBF{02:18:24.33 $-$05:22:56.8} &  \textBF{02:18:24.14 $-$05:22:55.3} & \textBF{16.7 $\pm$ 0.9} & \textBF{16.4 $\pm$ 1.3} & \textBF{\phantom{0}9.7 $\pm$ 0.7} & \textBF{\phantom{0}9.7 $\pm$ 0.7} \\ 
\phantom{\textBF{UDS}}\textBF{156.1}$^{\mathrm{b}}$ & \textBF{02:18:24.33 $-$05:22:56.8} &  \textBF{02:18:24.24 $-$05:22:56.9} & \textBF{16.7 $\pm$ 0.9} & \textBF{16.4 $\pm$ 1.3} & \textBF{\phantom{0}8.5 $\pm$ 0.7} & \textBF{\phantom{0}8.5 $\pm$ 0.7} \\ 
\textBF{UDS57.0}$^{\mathrm{b}}$ & \textBF{02:19:21.19 $-$04:56:52.5} &  \textBF{02:19:21.14 $-$04:56:51.3} & \textBF{13.0 $\pm$ 0.9} & \textBF{12.8 $\pm$ 1.7} & \textBF{\phantom{0}9.5 $\pm$ 0.6} & \textBF{\phantom{0}9.5 $\pm$ 0.6} \\ 
\phantom{\textBF{UDS}}\textBF{57.1}$^{\mathrm{b}}$ & \textBF{02:19:21.19 $-$04:56:52.5} &  \textBF{02:19:20.88 $-$04:56:52.9} & \textBF{13.0 $\pm$ 0.9} & \textBF{12.8 $\pm$ 1.7} & \textBF{\phantom{0}6.0 $\pm$ 0.9} & \textBF{\phantom{0}5.8 $\pm$ 0.9} \\ 
\phantom{\textBF{UDS}}\textBF{57.2}$^{\mathrm{b}}$ & \textBF{02:19:21.19 $-$04:56:52.5} &  \textBF{02:19:21.41 $-$04:56:49.0} & \textBF{13.0 $\pm$ 0.9} & \textBF{12.8 $\pm$ 1.7} & \textBF{\phantom{0}1.8 $\pm$ 0.6} & \textBF{\phantom{0}1.5 $\pm$ 0.6} \\ 
\phantom{\textBF{UDS}}\textBF{57.3}$^{\mathrm{b}}$ & \textBF{02:19:21.19 $-$04:56:52.5} &  \textBF{02:19:21.39 $-$04:56:38.8} & \textBF{13.0 $\pm$ 0.9} & \textBF{12.8 $\pm$ 1.7} & \textBF{\phantom{0}2.7 $\pm$ 1.0} & \textBF{\phantom{0}2.1 $\pm$ 1.0} \\ 
UDS03 & 02:15:55.41 $-$05:24:56.2 &  02:15:55.10 $-$05:24:56.6 & 12.8 $\pm$ 1.3 & 12.0 $\pm$ 1.8 & 13.7 $\pm$ 1.4 & \kern-1.1em$13.1^{\scriptscriptstyle +1.2}_{\scriptscriptstyle -1.5}$ \\ 
\textBF{UDS361.0}$^{\mathrm{b}}$ & \textBF{02:16:48.08 $-$05:01:30.7} &  \textBF{02:16:47.92 $-$05:01:29.8} & \textBF{11.5 $\pm$ 0.9} & \textBF{11.3 $\pm$ 1.7} & \textBF{11.8 $\pm$ 0.6} & \textBF{11.8 $\pm$ 0.6}\\ 
\phantom{\textBF{UDS}}\textBF{361.1}$^{\mathrm{b}}$ & \textBF{02:16:48.08 $-$05:01:30.7} &  \textBF{02:16:47.73 $-$05:01:25.8} & \textBF{11.5 $\pm$ 0.9} & \textBF{11.3 $\pm$ 1.7} & \textBF{\phantom{0}2.6 $\pm$ 0.7} & \textBF{\phantom{0}2.0 $\pm$ 0.7}\\  
\textBF{UDS286.0}$^{\mathrm{b}}$ & \textBF{02:17:25.81 $-$05:25:36.9} & \textBF{02:17:25.73 $-$05:25:41.2} & \textBF{11.4 $\pm$ 0.9} & \textBF{11.2 $\pm$ 1.7} & \textBF{\phantom{0}5.2 $\pm$ 0.7} & \textBF{\phantom{0}5.1 $\pm$ 0.7} \\ 
\phantom{\textBF{UDS}}\textBF{286.1}$^{\mathrm{b}}$ & \textBF{02:17:25.81 $-$05:25:36.9} &  \textBF{02:17:25.63 $-$05:25:33.7} & \textBF{11.4 $\pm$ 0.9} & \textBF{11.2 $\pm$ 1.7} & \textBF{\phantom{0}5.1 $\pm$ 0.6} & \textBF{\phantom{0}5.0 $\pm$ 0.6} \\ 
\phantom{\textBF{UDS}}\textBF{286.2}$^{\mathrm{b}}$ & \textBF{02:17:25.81 $-$05:25:36.9} &  \textBF{02:17:25.80 $-$05:25:37.5} & \textBF{11.4 $\pm$ 0.9} & \textBF{11.2 $\pm$ 1.7} & \textBF{\phantom{0}2.7 $\pm$ 0.6} & \textBF{\phantom{0}2.6 $\pm$ 0.6} \\ 
\phantom{\textBF{UDS}}\textBF{286.3}$^{\mathrm{b}}$ & \textBF{02:17:25.81 $-$05:25:36.9} &  \textBF{02:17:25.52 $-$05:25:36.7} & \textBF{11.4 $\pm$ 0.9} & \textBF{11.2 $\pm$ 1.7} & \textBF{\phantom{0}1.7 $\pm$ 0.6} & \textBF{\phantom{0}1.4 $\pm$ 0.6} \\ 
\textBF{UDS269.0}$^{\mathrm{b}}$ & \textBF{02:17:30.50 $-$05:19:22.9} & \textBF{02:17:30.44 $-$05:19:22.4} & \textBF{11.0 $\pm$ 0.9} & \textBF{10.7 $\pm$ 1.4} & \textBF{12.9 $\pm$ 0.6} & \textBF{12.9 $\pm$ 0.6} \\ 
\phantom{\textBF{UDS}}\textBF{269.1}$^{\mathrm{b}}$ & \textBF{02:17:30.50 $-$05:19:22.9} &  \textBF{02:17:30.25 $-$05:19:18.4} & \textBF{11.0 $\pm$ 0.9} & \textBF{10.7 $\pm$ 1.4} & \textBF{\phantom{0}2.6 $\pm$ 0.7} & \textBF{\phantom{0}2.1 $\pm$ 0.7} \\ 
UDS08 & 02:15:56.03 $-$04:55:10.3 &  02:15:55.95 $-$04:55:08.6 & 10.9 $\pm$ 1.0 & 10.5 $\pm$ 1.3 & 10.1 $\pm$ 1.7 & \kern-0.6em$8.9^{\scriptscriptstyle +1.6}_{\scriptscriptstyle -1.6}$ \\ 
\textBF{UDS204.0}$^{\mathrm{b}}$ & \textBF{02:18:03.04 $-$05:28:42.9} & \textBF{02:18:03.01 $-$05:28:41.9} & \textBF{10.7 $\pm$ 0.9} & \textBF{10.4 $\pm$ 1.2} & \textBF{11.6 $\pm$ 0.6} & \textBF{11.6 $\pm$ 0.6} \\ 
\phantom{\textBF{UDS}}\textBF{204.1}$^{\mathrm{b}}$ & \textBF{02:18:03.04 $-$05:28:42.9} &  \textBF{02:18:03.01 $-$05:28:32.5} & \textBF{10.7 $\pm$ 0.9} & \textBF{10.4 $\pm$ 1.2} & \textBF{\phantom{0}2.9 $\pm$ 0.9} & \textBF{\phantom{0}2.2 $\pm$ 0.9} \\ 
\textBF{UDS202.0}$^{\mathrm{b}}$ & \textBF{02:18:05.71 $-$05:10:50.9} & \textBF{02:18:05.65 $-$05:10:49.6} & \textBF{11.0 $\pm$ 0.9} & \textBF{10.4 $\pm$ 1.5} & \textBF{10.5 $\pm$ 0.5} & \textBF{10.5 $\pm$ 0.5} \\ 
\phantom{\textBF{UDS}}\textBF{202.1}$^{\mathrm{b}}$ & \textBF{02:18:05.71 $-$05:10:50.9} &  \textBF{02:18:05.05 $-$05:10:46.3} & \textBF{11.0 $\pm$ 0.9} & \textBF{10.4 $\pm$ 1.5} & \textBF{\phantom{0}3.9 $\pm$ 0.9} & \textBF{\phantom{0}3.5 $\pm$ 0.9} \\ 
UDS09 & 02:17:38.95 $-$04:33:37.0 &  02:17:38.82 $-$04:33:34.1 & 10.9 $\pm$ 1.3 & 10.1 $\pm$ 1.2 & 13.9 $\pm$ 0.8 & \kern-1.1em$13.6^{\scriptscriptstyle +0.9}_{\scriptscriptstyle -0.7}$ \\ 
UDS11 & 02:16:43.77 $-$05:17:54.7 &  02:16:43.72 $-$05:17:53.5 & 10.1 $\pm$ 0.9 & \phantom{0}9.8 $\pm$ 1.4 & 10.0 $\pm$ 1.8 & \kern-0.6em$8.6^{\scriptscriptstyle +1.7}_{\scriptscriptstyle -1.5}$  \\ 
\textBF{UDS306.0}$^{\mathrm{b}}$ & \textBF{02:17:17.23 $-$05:33:26.8} &  \textBF{02:17:17.07 $-$05:33:26.6} & \textBF{\phantom{0}9.9 $\pm$ 1.0} & \textBF{\phantom{0}9.7 $\pm$ 1.3} & \textBF{\phantom{0}8.3 $\pm$ 0.5} & \textBF{\phantom{0}8.3 $\pm$ 0.5} \\ 
\phantom{\textBF{UDS}}\textBF{306.1}$^{\mathrm{b}}$ & \textBF{02:17:17.23 $-$05:33:26.8} & \textBF{02:17:17.16 $-$05:33:32.5} & \textBF{\phantom{0}9.9 $\pm$ 1.0} & \textBF{\phantom{0}9.7 $\pm$ 1.3} & \textBF{\phantom{0}2.6 $\pm$ 0.4} & \textBF{\phantom{0}2.3 $\pm$ 0.4} \\ 
\phantom{\textBF{UDS}}\textBF{306.2}$^{\mathrm{b}}$ & \textBF{02:17:17.23 $-$05:33:26.8} &  \textBF{02:17:16.81 $-$05:33:31.8} & \textBF{\phantom{0}9.9 $\pm$ 1.0} & \textBF{\phantom{0}9.7 $\pm$ 1.3} & \textBF{\phantom{0}3.0 $\pm$ 0.9} & \textBF{\phantom{0}2.3 $\pm$ 0.9} \\ 
UDS14 & 02:16:30.77 $-$05:24:02.6 &  Undetected & \phantom{0}9.6 $\pm$ 0.9 & \phantom{0}9.4 $\pm$ 1.2 & ${<}\,6.1$ & {} \\ 
UDS15 & 02:18:03.57 $-$04:55:26.9 &  Undetected & \phantom{0}9.6 $\pm$ 0.9 & \phantom{0}9.4 $\pm$ 1.3 & ${<}\,5.1$ & {} \\ 
UDS16 & 02:19:02.24 $-$05:28:56.6 &  02:19:02.05 $-$05:28:56.7 & \phantom{0}9.5 $\pm$ 1.0 & \phantom{0}9.3 $\pm$ 1.4 & \phantom{0}6.5 $\pm$ 1.5 & \kern-0.6em$6.1^{\scriptscriptstyle +1.3}_{\scriptscriptstyle -1.6}$ \\ 
UDS18 & 02:17:44.29 $-$05:20:08.9 &  02:17:44.22 $-$05:20:09.8 & \phantom{0}9.3 $\pm$ 0.9 & \phantom{0}9.1 $\pm$ 1.3 & \phantom{0}8.9 $\pm$ 1.5 & \kern-0.6em$8.1^{\scriptscriptstyle +1.3}_{\scriptscriptstyle -1.4}$ \\ 
UDS13 & 02:19:27.31 $-$04:45:08.5 &  02:19:27.17 $-$04:45:06.1 & \phantom{0}9.8 $\pm$ 1.1 & \phantom{0}9.0 $\pm$ 1.6 & 15.3 $\pm$ 1.1 & \kern-1.1em$14.9^{\scriptscriptstyle +1.0}_{\scriptscriptstyle -1.2}$ \\ 
\textBF{UDS109.0}$^{\mathrm{b}}$ & \textBF{02:18:50.32 $-$05:27:22.7} &  \textBF{02:18:50.07 $-$05:27:25.5} & \textBF{\phantom{0}9.4 $\pm$ 0.9} & \textBF{\phantom{0}9.0 $\pm$ 1.5} & \textBF{\phantom{0}7.7 $\pm$ 0.7} & \textBF{\phantom{0}7.6 $\pm$ 0.7} \\ 
\phantom{\textBF{UDS}}\textBF{109.1}$^{\mathrm{b}}$ & \textBF{02:18:50.32 $-$05:27:22.7} &  \textBF{02:18:50.30 $-$05:27:17.2} & \textBF{\phantom{0}9.4 $\pm$ 0.9} & \textBF{\phantom{0}9.0 $\pm$ 1.5} & \textBF{\phantom{0}4.3 $\pm$ 0.6} & \textBF{\phantom{0}4.2 $\pm$ 0.6} \\ 
\textBF{UDS48.0}$^{\mathrm{b}}$ & \textBF{02:19:24.66 $-$04:53:00.5} &  \textBF{02:19:24.57 $-$04:53:00.2} & \textBF{\phantom{0}8.9 $\pm$ 0.8} & \textBF{\phantom{0}8.9 $\pm$ 1.3} & \textBF{\phantom{0}7.5 $\pm$ 0.5} & \textBF{\phantom{0}7.5 $\pm$ 0.5} \\ 
\phantom{\textBF{UDS}}\textBF{48.1}$^{\mathrm{b}}$ & \textBF{02:19:24.66 $-$04:53:00.5} &  \textBF{02:19:24.62 $-$04:52:56.9} & \textBF{\phantom{0}8.9 $\pm$ 0.8} & \textBF{\phantom{0}8.9 $\pm$ 1.3} & \textBF{\phantom{0}1.6 $\pm$ 0.5} & \textBF{\phantom{0}1.4 $\pm$ 0.5} \\ 
UDS20 & 02:17:30.51 $-$04:59:36.9 &  02:17:30.61 $-$04:59:36.8 & \phantom{0}9.1 $\pm$ 0.9 & \phantom{0}8.7 $\pm$ 1.4 & \phantom{0}9.0 $\pm$ 1.4 & \kern-0.6em$8.2^{\scriptscriptstyle +1.3}_{\scriptscriptstyle -1.3}$ \\ 
\textBF{UDS199.0}$^{\mathrm{b}}$ & \textBF{02:18:07.31 $-$04:44:12.9} &  \textBF{02:18:07.18 $-$04:44:13.8} & \textBF{\phantom{0}9.2 $\pm$ 0.9} & \textBF{\phantom{0}8.5 $\pm$ 1.4} & \textBF{\phantom{0}4.3 $\pm$ 0.6} & \textBF{\phantom{0}4.2 $\pm$ 0.6}\\ 
\phantom{\textBF{UDS}}\textBF{199.1}$^{\mathrm{b}}$ & \textBF{02:18:07.31 $-$04:44:12.9} &  \textBF{02:18:07.19 $-$04:44:10.9} & \textBF{\phantom{0}9.2 $\pm$ 0.9} & \textBF{\phantom{0}8.5 $\pm$ 1.4} & \textBF{\phantom{0}2.5 $\pm$ 0.5} & \textBF{\phantom{0}2.4 $\pm$ 0.5} \\ 
UDS22 & 02:16:11.81 $-$05:00:54.5 &  02:16:11.72 $-$05:00:54.0 & \phantom{0}9.0 $\pm$ 0.8 & \phantom{0}8.5 $\pm$ 1.2 & 15.0 $\pm$ 1.4 & \kern-1.1em$14.1^{\scriptscriptstyle +1.5}_{\scriptscriptstyle -1.3}$  \\ 
\textBF{UDS160.0}$^{\mathrm{b}}$ & \textBF{02:18:23.79 $-$05:11:40.9} &  \textBF{02:18:23.73 $-$05:11:38.5} & \textBF{\phantom{0}8.8 $\pm$ 0.9} & \textBF{\phantom{0}8.4 $\pm$ 1.4} & \textBF{\phantom{0}7.9 $\pm$ 0.6} & \textBF{\phantom{0}7.9 $\pm$ 0.6}\\ 
\textBF{UDS110.0}$^{\mathrm{b}}$ & \textBF{02:18:48.43 $-$05:18:06.7} &  \textBF{02:18:48.24 $-$05:18:05.2} & \textBF{\phantom{0}8.4 $\pm$ 0.9} & \textBF{\phantom{0}8.2 $\pm$ 1.4} & \textBF{\phantom{0}7.7 $\pm$ 0.6} & \textBF{\phantom{0}7.7 $\pm$ 0.6} \\ 
\phantom{\textBF{UDS}}\textBF{110.1}$^{\mathrm{b}}$ & \textBF{02:18:48.43 $-$05:18:06.7} &  \textBF{02:18:48.76 $-$05:18:02.1} & \textBF{\phantom{0}8.4 $\pm$ 0.9} & \textBF{\phantom{0}8.2 $\pm$ 1.4} & \textBF{\phantom{0}2.5 $\pm$ 0.8} & \textBF{\phantom{0}2.0 $\pm$ 0.8} \\ 
UDS21 & 02:19:34.14 $-$04:44:40.4 &  02:19:34.15 $-$04:44:38.1 & \phantom{0}9.0 $\pm$ 1.2 & \phantom{0}8.2 $\pm$ 1.5 & 10.3 $\pm$ 1.0 & \kern-0.6em$9.9^{\scriptscriptstyle +0.9}_{\scriptscriptstyle -1.0}$  \\ 
\textBF{UDS337.0}$^{\mathrm{b}}$ & \textBF{02:16:41.11 $-$05:03:52.7} &  \textBF{02:16:41.11 $-$05:03:51.4} & \textBF{\phantom{0}8.4 $\pm$ 0.9} & \textBF{\phantom{0}8.0 $\pm$ 1.2} & \textBF{\phantom{0}8.1 $\pm$ 0.5} & \textBF{\phantom{0}8.1 $\pm$ 0.5} \\ 
UDS29 & 02:17:39.87 $-$05:29:18.9 &  02:17:39.78 $-$05:29:19.1 & \phantom{0}8.3 $\pm$ 0.9 & \phantom{0}8.0 $\pm$ 1.3 & 11.6 $\pm$ 1.1 & \kern-1.1em$11.2^{\scriptscriptstyle +1.0}_{\scriptscriptstyle -1.2}$ \\ 
\textBF{UDS79.0}$^{\mathrm{b}}$ & \textBF{02:19:10.09 $-$05:00:08.6} &  \textBF{02:19:09.94 $-$05:00:08.6} & \textBF{\phantom{0}8.1 $\pm$ 0.9} & \textBF{\phantom{0}7.9 $\pm$ 1.4} & \textBF{\phantom{0}7.7 $\pm$ 0.5} & \textBF{\phantom{0}7.7 $\pm$ 0.5} \\ 
UDS30 & 02:17:55.27 $-$04:47:22.9 &  02:17:55.05 $-$04:47:22.9 & \phantom{0}8.3 $\pm$ 0.9 & \phantom{0}7.8 $\pm$ 1.2 & \phantom{0}7.4 $\pm$ 1.1 & \kern-0.6em$7.1^{\scriptscriptstyle +1.0}_{\scriptscriptstyle -1.0}$ \\ 
UDS28 & 02:19:42.53 $-$05:18:04.3 &  02:19:42.45 $-$05:18:03.6 & \phantom{0}8.4 $\pm$ 1.1 & \phantom{0}7.6 $\pm$ 1.6 & \phantom{0}9.0 $\pm$ 1.0 & \kern-0.6em$8.6^{\scriptscriptstyle +0.9}_{\scriptscriptstyle -1.0}$ \\ 
UDS36 & 02:17:12.19 $-$04:43:18.9 &  02:17:12.21 $-$04:43:16.5 & \phantom{0}8.0 $\pm$ 0.9 & \phantom{0}7.6 $\pm$ 1.2 & \phantom{0}8.5 $\pm$ 1.4 & \kern-0.6em$7.8^{\scriptscriptstyle +1.3}_{\scriptscriptstyle -1.2}$ \\ 
UDS34 & 02:17:42.15 $-$04:56:28.9 &  02:17:41.92 $-$04:56:29.8 & \phantom{0}8.0 $\pm$ 0.9 & \phantom{0}7.6 $\pm$ 1.3 & \phantom{0}7.9 $\pm$ 1.2 & \kern-0.6em$7.6^{\scriptscriptstyle +1.0}_{\scriptscriptstyle -1.3}$ \\ 
UDS35 & 02:16:40.43 $-$05:13:38.7 &  02:16:40.40 $-$05:13:35.9 & \phantom{0}8.0 $\pm$ 0.9 & \phantom{0}7.6 $\pm$ 1.3 & \phantom{0}7.1 $\pm$ 1.4 & \kern-0.6em$6.6^{\scriptscriptstyle +1.3}_{\scriptscriptstyle -1.4}$ \\ 
UDS37 & 02:16:38.44 $-$05:01:22.7 &  02:16:38.33 $-$05:01:21.4 & \phantom{0}7.9 $\pm$ 0.9 & \phantom{0}7.5 $\pm$ 1.3 & \phantom{0}8.4 $\pm$ 1.3 & \kern-0.6em$7.8^{\scriptscriptstyle +1.2}_{\scriptscriptstyle -1.2}$ \\ 
UDS39 & 02:16:40.57 $-$05:11:00.7 &  02:16:40.59 $-$05:10:58.8 & \phantom{0}7.9 $\pm$ 0.9 & \phantom{0}7.5 $\pm$ 1.4 & \phantom{0}7.9 $\pm$ 1.0 & \kern-0.6em$7.6^{\scriptscriptstyle +0.9}_{\scriptscriptstyle -1.0}$ \\ 
UDS40 & 02:17:27.43 $-$05:06:44.9 &  02:17:27.29 $-$05:06:42.8 & \phantom{0}7.8 $\pm$ 0.9 & \phantom{0}7.5 $\pm$ 1.2 & \phantom{0}6.9 $\pm$ 1.1 & \kern-0.6em$6.6^{\scriptscriptstyle +1.1}_{\scriptscriptstyle -1.0}$ \\ 
\textBF{UDS168.0}$^{\mathrm{b}}$ & \textBF{02:18:20.46 $-$05:31:44.8} &  \textBF{02:18:20.40 $-$05:31:43.2} & \textBF{\phantom{0}8.2 $\pm$ 0.9} & \textBF{\phantom{0}7.5 $\pm$ 1.4} & \textBF{\phantom{0}6.7 $\pm$ 0.6} & \textBF{\phantom{0}6.7 $\pm$ 0.6} \\ 
\phantom{\textBF{UDS}}\textBF{168.1}$^{\mathrm{b}}$ & \textBF{02:18:20.46 $-$05:31:44.8} &  \textBF{02:18:20.31 $-$05:31:41.7} & \textBF{\phantom{0}8.2 $\pm$ 0.9} & \textBF{\phantom{0}7.5 $\pm$ 1.4} & \textBF{\phantom{0}3.0 $\pm$ 0.6} & \textBF{\phantom{0}2.8 $\pm$ 0.6}\\ 
\phantom{\textBF{UDS}}\textBF{168.2}$^{\mathrm{b}}$ & \textBF{02:18:20.46 $-$05:31:44.8} &  \textBF{02:18:20.17 $-$05:31:38.6} & \textBF{\phantom{0}8.2 $\pm$ 0.9} & \textBF{\phantom{0}7.5 $\pm$ 1.4} & \textBF{\phantom{0}2.0 $\pm$ 0.7} & \textBF{\phantom{0}1.6 $\pm$ 0.7} \\ 
UDS33 & 02:15:46.99 $-$05:18:52.2 &  02:15:46.70 $-$05:18:49.2 & \phantom{0}8.1 $\pm$ 1.2 & \phantom{0}7.4 $\pm$ 1.4 & 10.3 $\pm$ 1.0 & \kern-0.6em$9.9^{\scriptscriptstyle +1.0}_{\scriptscriptstyle -1.0}$ \\ 
\textBF{UDS218.0}$^{\mathrm{b}}$ & \textBF{02:17:54.87 $-$05:23:22.9} & \textBF{02:17:54.80 $-$05:23:23.0} & \textBF{\phantom{0}7.6 $\pm$ 0.9} & \textBF{\phantom{0}7.2 $\pm$ 1.3} & \textBF{\phantom{0}6.6 $\pm$ 0.7} & \textBF{\phantom{0}6.6 $\pm$ 0.7} \\ 
UDS38 & 02:16:46.07 $-$05:03:46.7 &  02:16:46.17 $-$05:03:48.9 & \phantom{0}7.9 $\pm$ 0.9 & \phantom{0}7.2 $\pm$ 1.3 & \phantom{0}6.9 $\pm$ 1.6 & \kern-0.6em$6.3^{\scriptscriptstyle +1.5}_{\scriptscriptstyle -1.5}$ \\ 
\hline
\end{tabular}
\begin{tablenotes}
\small 
\item $^{\mathrm{a}}$ From \citet{ikarashi2011} using the SMA at 860\,$\mu$m. 
\item $^{\mathrm{b}}$ From \citet{simpson2015} using ALMA at 870\,$\mu$m, following the naming convention in their paper.
\end{tablenotes}
\end{threeparttable}
\end{table*}

\begin{table*}
\caption{SMA sample for the SSA22 field, ordered by decreasing deboosted SCUBA-2 flux density. The source observed by the SMA in \citet{tamura2010} is in bold and is indicated by a $^{\mathrm{c}}$, and all other sources were observed by the SMA in this work.}
\label{table2}
\begin{threeparttable}
\begin{tabular}{lcccccccc|}
\hline                                 
Source & RA/Dec SCUBA-2 & RA/Dec SMA & $S^{\rm obs}_{\rm S2}$ [mJy] & $S_{\rm S2}$ [mJy] & $S^{\rm obs}_{\rm SMA}$ [mJy] &  $S_{\rm SMA}$ [mJy] \\
{} & (J2000) & (J2000) & {} & {} & {} & {} \\
 \hline
\textBF{SSA22-AzTEC1}$^{\mathrm{c}}$ & \textBF{22:17:32.50 $+$00:17:40.4} &  \textBF{22:17:32.42 $+$00:17:44.0} & \textBF{14.5 $\pm$ 1.1} & \textBF{14.5 $\pm$ 1.4} & \textBF{12.2 $\pm$ 2.3} & {} \\ 
 SSA22-03 & 22:16:56.10 $+$00:28:44.4 &  Undetected & 11.1 $\pm$ 1.2 & 10.7 $\pm$ 1.4 & ${<}\,8.7$ & {} \\ 
SSA22-02 & 22:16:59.96 $+$00:10:40.4 &  22:16:59.83 $+$00:10:37.1 & 10.8 $\pm$ 1.1 & 10.2 $\pm$ 1.5 & \phantom{0}9.3 $\pm$ 1.6 & \phantom{0}$8.2^{\scriptscriptstyle +1.5}_{\scriptscriptstyle -1.6}$ \\ 
SSA22-04 & 22:16:51.43 $+$00:18:20.4 &  Undetected & 10.4 $\pm$ 1.1 & 10.0 $\pm$ 1.4 & ${<}\,12.6$ & {} \\ 
SSA22-08 & 22:18:06.63 $+$00:05:20.4 &  22:18:06.60 $+$00:05:20.5 & 10.0 $\pm$ 1.3 & \phantom{0}8.8 $\pm$ 1.8 & \phantom{0}9.5 $\pm$ 1.6 & \phantom{0}$8.2^{\scriptscriptstyle +1.7}_{\scriptscriptstyle -1.3}$ \\ 
SSA22-07 & 22:17:18.90 $+$00:18:06.4 &  Undetected & \phantom{0}8.5 $\pm$ 1.1 & \phantom{0}7.9 $\pm$ 1.3 & ${<}\,8.5$ & {} \\ 
SSA22-06 & 22:18:06.36 $+$00:11:34.4 &  22:18:06.48 $+$00:11:34.7 & \phantom{0}8.3 $\pm$ 1.1 & \phantom{0}7.7 $\pm$ 1.5 & \phantom{0}9.9 $\pm$ 1.3 & \phantom{0}$9.2^{\scriptscriptstyle +1.3}_{\scriptscriptstyle -1.3}$ \\ 
SSA22-05 & 22:17:34.10 $+$00:13:52.4 &  22:17:33.90 $+$00:13:52.3 & \phantom{0}7.9 $\pm$ 1.1 & \phantom{0}7.3 $\pm$ 1.1 & 11.7 $\pm$ 2.0 & \phantom{0}$9.9^{\scriptscriptstyle +2.0}_{\scriptscriptstyle -1.8}$ \\ 
SSA22-09 & 22:17:42.23 $+$00:17:00.4 &  Undetected & \phantom{0}6.7 $\pm$ 1.1 & \phantom{0}6.0 $\pm$ 1.4 & ${<}\,8.5$ & {}\\ 
 \hline
\end{tabular}
\begin{tablenotes}
\small 
\item $^{\mathrm{c}}$ From \citet{tamura2010} using the SMA at 860\,$\mu$m, following the naming convention in their paper. 
\end{tablenotes}
\end{threeparttable}
\end{table*}

\begin{table*}
\caption{SMA sample plus archival SMA data for the COSMOS field, ordered by decreasing deboosted SCUBA-2 flux density. Sources observed by the SMA in \citet{younger2007} are in bold and indicated by a $^{\mathrm{d}}$, sources observed by the SMA in \citet{younger2009} are in bold and indicated by a $^{\mathrm{e}}$, the source observed by the SMA in \citet{aravena2010} is in bold and indicated by a $^{\mathrm{f}}$, the sources observed by CARMA and PdBI in \citet{smolcic2012b} and \citet{smolcic2012} are in bold and indicated by a $^{\mathrm{g}}$ and a $^{\mathrm{h}}$, respectively, and all other sources were observed by the SMA in this work. Flux density measurements from \citet{younger2007,younger2009}, \citet{aravena2010} and \citet{smolcic2012b,smolcic2012} were not deboosted. Values of N/A in the $S_{\rm SMA}$ column indicate sources where our deboosting simulation was not applicable.}
\label{table3}
\begin{threeparttable}
\begin{tabular}{lccccccc|}
\hline                                 
Source & RA/Dec SCUBA-2 & RA/Dec SMA & $S^{\rm obs}_{\rm S2}$ [mJy] & $S_{\rm S2}$ [mJy] & $S^{\rm obs}_{\rm SMA}$ [mJy] &  $S_{\rm SMA}$ [mJy] \\
{} & (J2000) & (J2000) & {} & {} & {} & {} \\
 \hline
 \textBF{AzTEC1}$^{\mathrm{d}}$ & \textBF{09:59:42.89 $+$02:29:36.5} &  \textBF{09:59:42.86 $+$02:29:38.2} & \textBF{16.7 $\pm$ 1.5} & \textBF{16.0 $\pm$ 3.0} & \textBF{15.6 $\pm$ 1.1} & {} \\ 
\textBF{AzTEC2}$^{\mathrm{d}}$ & \textBF{10:00:08.11 $+$02:26:12.6} &  \textBF{10:00:08.05 $+$02:26:12.2} & \textBF{15.4 $\pm$ 1.4} & \textBF{14.7 $\pm$ 2.3} & \textBF{12.4 $\pm$ 1.0} & {} \\ 
COSMOS05 & 09:59:22.99 $+$02:51:36.4 &  09:59:22.99 $+$02:51:36.4 & 14.0 $\pm$ 1.5 & 13.0 $\pm$ 1.7 & 13.7 $\pm$ 2.3 & $11.3^{\scriptscriptstyle +2.4}_{\scriptscriptstyle -2.2}$ \\ 
COSMOS06 & 09:58:42.40 $+$02:54:42.2 &  Undetected & 14.0 $\pm$ 1.5 & 13.0 $\pm$ 2.1 & ${<}\,8.1$ & {} \\ 
\textBF{MM1}$^{\mathrm{f}}$ & \textBF{10:00:15.72 $+$02:15:48.6} &  \textBF{10:00:15.61 $+$02:15:49.0} & \textBF{12.9 $\pm$ 0.8} & \textBF{12.9 $\pm$ 1.2} & \textBF{16.8 $\pm$ 1.5} & {} \\ 
\textBF{COSLA-54}$^{\mathrm{h}}$ & \textBF{09:58:37.92 $+$02:14:06.3} &  \textBF{09:58:37.99 $+$02:14:08.5} & \textBF{13.2 $\pm$ 1.0} & \textBF{12.4 $\pm$ 1.5} & \textBF{12.7 $\pm$ 2.5} & {} \\ 
\textBF{Cosbo-3}$^{\mathrm{g}}$ & \textBF{10:00:57.22 $+$02:20:12.6} &  \textBF{10:00:56.95 $+$02:20:17.8} & \textBF{13.0 $\pm$ 1.5} & \textBF{12.1 $\pm$ 2.2} & \textBF{10.9 $\pm$ 2.7} & {} \\ 
\textBF{AzTEC9}$^{\mathrm{e}}$ & \textBF{09:59:57.44 $+$02:27:28.6} & \textBF{09:59:57.25 $+$02:27:30.6} & \textBF{12.4 $\pm$ 1.4} & \textBF{11.8 $\pm$ 1.9} & \textBF{\phantom{0}9.0 $\pm$ 2.2} & {} \\ 
COSMOS08 & 09:59:10.31 $+$02:48:54.4 &  09:59:10.34 $+$02:48:55.5 & 13.1 $\pm$ 1.6 & 11.7 $\pm$ 2.1 & 12.7 $\pm$ 2.0 & $11.3^{\scriptscriptstyle +1.6}_{\scriptscriptstyle -2.3}$ \\ 
COSMOS11a & 09:58:45.89 $+$02:43:26.3 &  09:58:45.95 $+$02:43:29.1 & 12.5 $\pm$ 1.6 & 11.5 $\pm$ 2.0 & \phantom{0}8.6 $\pm$ 1.1 & \phantom{0}$8.0^{\scriptscriptstyle +1.1}_{\scriptscriptstyle -1.0}$ \\ 
\phantom{COSMOS}11b & 09:58:45.89 $+$02:43:26.3 &  09:58:46.06 $+$02:43:31.5 & 12.5 $\pm$ 1.6 & 11.5 $\pm$ 2.0 & \phantom{0}5.1 $\pm$ 1.1 & N/A \\ 
COSMOS15 & 09:57:49.03 $+$02:46:15.9 &  09:57:48.93 $+$02:46:19.9 & 11.8 $\pm$ 1.5 & 11.2 $\pm$ 2.1 & 11.2 $\pm$ 2.0 & \phantom{0}$9.7^{\scriptscriptstyle +1.6}_{\scriptscriptstyle -2.2}$ \\ 
\textBF{AzTEC5}$^{\mathrm{d}}$ & \textBF{10:00:19.86 $+$02:32:04.6} &  \textBF{10:00:19.75 $+$02:32:04.4} & \textBF{12.0 $\pm$ 1.4} & \textBF{11.2 $\pm$ 2.2} & \textBF{\phantom{0}9.3 $\pm$ 1.3} & {} \\ 
COSMOS14 & 10:00:13.46 $+$01:37:04.7 &  10:00:13.47 $+$01:37:04.3 & 12.0 $\pm$ 1.5 & 11.0 $\pm$ 1.8 & 12.2 $\pm$ 1.2 & $11.7^{\scriptscriptstyle +1.0}_{\scriptscriptstyle -1.3}$ \\ 
COSMOS17 & 10:00:04.78 $+$02:30:44.6 &  Undetected & 11.2 $\pm$ 1.4 & 11.0 $\pm$ 1.8 & ${<}\,8.4$ & {} \\ 
\textBF{AzTEC12}$^{\mathrm{e}}$ & \textBF{10:00:35.34 $+$02:43:52.6} &  \textBF{10:00:35.29 $+$02:43:53.4} & \textBF{11.6 $\pm$ 1.3} & \textBF{10.9 $\pm$ 2.0} & \textBF{13.5 $\pm$ 1.8} & {} \\ 
COSMOS18 & 09:58:40.46 $+$02:05:14.4 &  09:58:40.28 $+$02:05:14.5 & 11.1 $\pm$ 1.5 & 10.4 $\pm$ 2.1 & 10.9 $\pm$ 1.7 & \phantom{0}$9.7^{\scriptscriptstyle +1.6}_{\scriptscriptstyle -1.7}$ \\
\textBF{AzTEC8}$^{\mathrm{e}}$ & \textBF{09:59:59.44 $+$02:34:38.6} & \textBF{09:59:59.34 $+$02:34:41.0} & \textBF{10.9 $\pm$ 1.4} & \textBF{10.1 $\pm$ 1.8} & \textBF{19.7 $\pm$ 1.8} & {} \\ 
\textBF{AzTEC7}$^{\mathrm{d}}$ & \textBF{10:00:17.99 $+$02:48:30.5} &  \textBF{10:00:18.06 $+$02:48:30.5} & \textBF{10.8 $\pm$ 1.4} & \textBF{\phantom{0}9.7 $\pm$ 2.0} & \textBF{12.0 $\pm$ 1.5} & {} \\ 
COSMOS21 & 09:59:07.63 $+$02:58:36.3 &  09:59:07.49 $+$02:58:39.3 & 10.6 $\pm$ 1.5 & \phantom{0}9.5 $\pm$ 2.0 & \phantom{0}9.9 $\pm$ 1.9 & \phantom{0}$8.3^{\scriptscriptstyle +1.8}_{\scriptscriptstyle -1.8}$ \\ 
\textBF{AzTEC3}$^{\mathrm{d}}$ & \textBF{10:00:20.79 $+$02:35:20.6} &  \textBF{10:00:20.70 $+$02:35:20.5} & \textBF{\phantom{0}9.2 $\pm$ 1.3} & \textBF{\phantom{0}8.6 $\pm$ 1.5} & \textBF{\phantom{0}8.7 $\pm$ 1.5} & {} \\ 
\textBF{COSLA-13}$^{\mathrm{h}}$ & \textBF{10:00:31.87 $+$02:12:42.6} &  \textBF{10:00:31.84 $+$02:12:42.8} & \textBF{\phantom{0}9.1 $\pm$ 0.9} & \textBF{\phantom{0}8.4 $\pm$ 1.4} & \textBF{9.3 $\pm$ 2.4} & {} \\ 
\textBF{AzTEC11.N}$^{\mathrm{e}}$ & \textBF{10:00:08.91 $+$02:40:10.6} &  \textBF{10:00:08.91 $+$02:40:09.6} & \textBF{\phantom{0}9.3 $\pm$ 1.4} & \textBF{\phantom{0}8.3 $\pm$ 1.8} & \textBF{10.0 $\pm$ 2.1} & {} \\ 
\phantom{\textBF{AzTEC}}\textBF{11.S}$^{\mathrm{e}}$ & \textBF{10:00:08.91 $+$02:40:10.6} &  \textBF{10:00:08.94 $+$02:40:12.3} & \textBF{\phantom{0}9.3 $\pm$ 1.4} & \textBF{\phantom{0}8.3 $\pm$ 1.8} & \textBF{\phantom{0}4.4 $\pm$ 2.1} & {} \\ 
\textBF{COSLA-23-N}$^{\mathrm{h}}$ & \textBF{10:00:10.12 $+$02:13:34.6} &  \textBF{10:00:10.16 $+$02:13:35.0} & \textBF{\phantom{0}8.4 $\pm$ 0.9} & \textBF{\phantom{0}8.2 $\pm$ 1.4} & \textBF{13.4 $\pm$ 1.8} & {} \\ 
\textBF{\phantom{COSLA-}23-S}$^{\mathrm{h}}$ & \textBF{10:00:10.12 $+$02:13:34.6} &  \textBF{10:00:10.07 $+$02:13:26.9} & \textBF{\phantom{0}8.4 $\pm$ 0.9} & \textBF{\phantom{0}8.2 $\pm$ 1.4} & \textBF{14.5 $\pm$ 2.3} & {} \\ 
\textBF{AzTEC6}$^{\mathrm{d}}$ & \textBF{10:00:06.64 $+$02:38:34.6} &  \textBF{10:00:06.50 $+$02:38:37.7} & \textBF{\phantom{0}8.9 $\pm$ 1.4} & \textBF{\phantom{0}8.0 $\pm$ 1.8} & \textBF{\phantom{0}8.6 $\pm$ 1.3} & {} \\ 
\textBF{AzTEC4}$^{\mathrm{d}}$ & \textBF{09:59:31.68 $+$02:30:42.5} & \textBF{09:59:31.72 $+$02:30:44.0} & \textBF{\phantom{0}9.3 $\pm$ 1.5} & \textBF{\phantom{0}7.9 $\pm$ 1.9} & \textBF{14.4 $\pm$ 1.9} & {} \\ 
COSMOS22 & 09:59:33.55 $+$02:23:46.5 &  09:59:33.55 $+$02:23:46.5 & \phantom{0}8.5 $\pm$ 1.2 & \phantom{0}7.8 $\pm$ 1.6 & \phantom{0}8.9 $\pm$ 2.2 & ${<}\,8.9$ \\ 
\textBF{COSLA-35}$^{\mathrm{h}}$ & \textBF{10:00:23.59 $+$02:21:54.6}  &  \textBF{10:00:23.65 $+$02:21:55.2} & \textBF{\phantom{0}8.0 $\pm$ 1.1} & \textBF{\phantom{0}7.3 $\pm$ 1.5} & \textBF{\phantom{0}8.4 $\pm$ 2.0} & {} \\ 
COSMOS24 & 09:59:12.08 $+$02:09:54.5  &  09:59:12.17 $+$02:09:57.1 & \phantom{0}7.9 $\pm$ 1.1 & \phantom{0}7.2 $\pm$ 1.3 & \phantom{0}8.1 $\pm$ 1.7 & \phantom{0}$7.0^{\scriptscriptstyle +1.6}_{\scriptscriptstyle -1.6}$ \\ 
COSMOS25 & 10:00:23.73 $+$02:19:14.6 &  Undetected & \phantom{0}7.2 $\pm$ 1.0 & \phantom{0}7.1 $\pm$ 1.1 & ${<}\,9.3$ & {} \\ 
\hline
 COSMOS01$^{\mathrm{i}}$ & 10:02:09.77 $+$02:36:33.9 &  10:02:09.64 $+$02:36:32.5 & 20.3 $\pm$ 3.3 & {} & 10.6 $\pm$ 1.2 & {} \\ 
 COSMOS02$^{\mathrm{i}}$ & 10:02:49.22 $+$02:32:55.1 &  10:02:49.19 $+$02:32:55.3 & 20.2 $\pm$ 3.6 & {} & 18.6 $\pm$ 0.7 & {} \\ 
\hline
\end{tabular}
\begin{tablenotes}
\small
\item $^{\mathrm{d}}$ From \citet{younger2007} using the SMA at 890\,$\mu$m, following the naming convention in their paper. 
\item $^{\mathrm{e}}$ From \citet{younger2009} using the SMA at 890\,$\mu$m, following the naming convention in their paper.
\item $^{\mathrm{f}}$ From \citet{aravena2010} using the SMA at 890\,$\mu$m, following the naming convention in their paper.
\item $^{\mathrm{g}}$ From \citet{smolcic2012b} using CARMA at 1.3\,mm and extrapolated to 860\,$\mu$m using a modified blackbody with dust a temperature of 35\,K, a dust spectral index of 2 and a redshift of 2, following the naming convention in their paper.
\item $^{\mathrm{h}}$ From \citet{smolcic2012} using PdBI at 1.3\,mm and extrapolated to 860\,$\mu$m using a modified blackbody with dust a temperature of 35\,K, a dust spectral index of 2 and a redshift of 2, following the naming convention in their paper.
\item $^{\mathrm{i}}$ Source is found in the S2CLS maps but outside the area defining the S2CLS catalogue, and hence not used in our analysis. 
\end{tablenotes}
\end{threeparttable}
\end{table*}

\begin{table*}
\caption{SMA sample for the LHN field, ordered by decreasing deboosted SCUBA-2 flux density. All observations are from this work.  Values of N/A in the $S_{\rm SMA}$ column indicate sources where our deboosting simulation was not applicable.}
\label{table4}
\begin{threeparttable}
\begin{tabular}{lccccccc|}
\hline                                 
Source & RA/Dec SCUBA-2 & RA/Dec SMA & $S^{\rm obs}_{\rm S2}$ [mJy] & $S_{\rm S2}$ [mJy] & $S^{\rm obs}_{\rm SMA}$ [mJy] &  $S_{\rm SMA}$ [mJy] \\
{} & (J2000) & (J2000) & {} & {} & {} & {} \\
 \hline
LHN01 & 10:46:45.01 $+$59:15:39.8 &  10:46:45.00 $+$59:15:41.6 & 12.3 $\pm$ 1.2 &12.3 $\pm$ 1.8 & 10.3 $\pm$ 1.9 & \phantom{0}$8.8^{\scriptscriptstyle +1.8}_{\scriptscriptstyle -1.7}$ \\ 
LHN02 & 10:46:35.78 $+$59:07:48.0 & 10:46:35.91 $+$59:07:48.1 & 12.0 $\pm$ 1.0 & 11.9 $\pm$ 1.2 & 12.2 $\pm$ 1.9 & $10.4^{\scriptscriptstyle +2.0}_{\scriptscriptstyle -1.7}$ \\ 
LHN03a & 10:47:27.66 $+$58:52:14.6 &  10:47:27.97 $+$58:52:14.1 & 10.4 $\pm$ 1.1 & \phantom{0}9.9 $\pm$ 1.3 & \phantom{0}8.1 $\pm$ 1.8 & \phantom{0}$7.3^{\scriptscriptstyle +1.5}_{\scriptscriptstyle -1.8}$ \\ 
\phantom{LHN}03b & 10:47:27.66 $+$58:52:14.6 &  10:47:26.52 $+$58:52:12.8 & 10.4 $\pm$ 1.1 & \phantom{0}9.9 $\pm$ 1.3 & \phantom{0}8.0 $\pm$ 1.9 & \phantom{0}$7.1^{\scriptscriptstyle +1.6}_{\scriptscriptstyle -1.9}$ \\ 
LHN06 & 10:45:55.19 $+$59:15:28.1 &  10:45:55.24 $+$59:15:28.6 & \phantom{0}9.7 $\pm$ 1.1 & \phantom{0}9.7 $\pm$ 0.9 & \phantom{0}7.2 $\pm$ 1.8 & \phantom{0}$6.6^{\scriptscriptstyle +1.5}_{\scriptscriptstyle -6.5}$ \\ 
LHN04 & 10:48:03.37 $+$58:54:22.9 &  10:48:03.57 $+$58:54:21.5 & 10.1 $\pm$ 1.3 & \phantom{0}8.9 $\pm$ 1.4 & 14.1 $\pm$ 2.4 & $11.7^{+2.2}_{-2.5}$ \\ 
LHN08 & 10:47:00.03 $+$59:01:07.5 &  10:47:00.18 $+$59:01:07.5 & \phantom{0}9.2 $\pm$ 1.0 & \phantom{0}8.9 $\pm$ 1.6 & 10.4 $\pm$ 1.6 & \phantom{0}$9.4^{\scriptscriptstyle +1.4}_{\scriptscriptstyle -1.6}$ \\ 
LHN11 & 10:45:22.55 $+$59:17:21.7 &  Undetected & \phantom{0}8.6 $\pm$ 1.4 & \phantom{0}8.8 $\pm$ 1.7 & ${<}\,7.2$ & {} \\
LHN07 & 10:45:35.23 $+$58:50:49.9 &  10:45:34.98 $+$58:50:49.9 & \phantom{0}9.3 $\pm$ 1.1 & \phantom{0}8.7 $\pm$ 1.4 & \phantom{0}9.6 $\pm$ 1.6 & \phantom{0}$8.8^{\scriptscriptstyle +1.3}_{\scriptscriptstyle -1.6}$ \\ 
LHN10 & 10:45:54.58 $+$58:47:54.1 &  10:45:54.50 $+$58:47:55.6 & \phantom{0}8.8 $\pm$ 1.1 & \phantom{0}8.3 $\pm$ 1.5 & \phantom{0}8.2 $\pm$ 0.8 & \phantom{0}$8.1^{\scriptscriptstyle +0.7}_{\scriptscriptstyle -0.8}$ \\ 
LHN05 & 10:43:51.48 $+$59:00:57.7 &  10:43:51.21 $+$59:00:58.1 & 10.0 $\pm$ 1.5 & \phantom{0}8.2 $\pm$ 2.1 & 10.9 $\pm$ 2.4 & \phantom{0}$8.8^{\scriptscriptstyle +2.0}_{\scriptscriptstyle -2.3}$ \\ 
LHN09 & 10:45:23.87 $+$59:16:25.7 &  10:45:23.11 $+$59:16:18.6 & \phantom{0}9.0 $\pm$ 1.3 & \phantom{0}8.2 $\pm$ 1.5 & \phantom{0}9.4 $\pm$ 1.5 & \phantom{0}$8.6^{\scriptscriptstyle +1.3}_{\scriptscriptstyle -1.4}$ \\ 
LHN12 & 10:46:32.85 $+$59:02:12.0 &  Undetected & \phantom{0}8.6 $\pm$ 1.0 & \phantom{0}8.1 $\pm$ 1.3 & ${<}\,8.0$ & {} \\ 
LHN13a & 10:47:25.25 $+$59:03:40.7 & 10:47:25.47 $+$59:03:36.7 & \phantom{0}8.5 $\pm$ 1.1 & \phantom{0}7.9 $\pm$ 1.4 & \phantom{0}5.5 $\pm$ 0.8 & N/A \\ 
\phantom{LHN}13b & 10:47:25.25 $+$59:03:40.7 &  10:47:25.13 $+$59:03:41.5 & \phantom{0}8.5 $\pm$ 1.1 & \phantom{0}7.9 $\pm$ 1.4 & \phantom{0}3.9 $\pm$ 0.8 & N/A \\ 
LHN14 & 10:46:31.68 $+$58:50:54.0 & 10:46:31.58 $+$58:50:55.7 & \phantom{0}8.5 $\pm$ 1.1 & \phantom{0}7.9 $\pm$ 1.4 & \phantom{0}7.1 $\pm$ 0.8 & \phantom{0}$7.0^{\scriptscriptstyle +0.7}_{\scriptscriptstyle -0.8}$ \\ 
LHN15 & 10:46:57.26 $+$59:14:57.6 &  10:46:57.30 $+$59:14:58.6 & \phantom{0}8.5 $\pm$ 1.2 & \phantom{0}7.9 $\pm$ 0.9 & \phantom{0}5.5 $\pm$ 0.7 & \phantom{0}$5.5^{\scriptscriptstyle +0.6}_{\scriptscriptstyle -0.8}$ \\ 
LHN16 & 10:44:56.86 $+$58:49:59.0 &  10:44:56.74 $+$58:49:59.7 & \phantom{0}8.3 $\pm$ 1.1 & \phantom{0}7.6 $\pm$ 1.4 & 16.9 $\pm$ 2.5 & $13.9^{\scriptscriptstyle +3.0}_{\scriptscriptstyle -2.2}$ \\ 
LHN17 & 10:44:47.69 $+$59:00:36.6 & 10:44:47.68 $+$59:00:35.6 & \phantom{0}8.1 $\pm$ 1.1 & \phantom{0}7.5 $\pm$ 1.3 & \phantom{0}5.6 $\pm$ 0.7 & \phantom{0}$5.5^{\scriptscriptstyle +0.7}_{\scriptscriptstyle -0.7}$ \\ 
LHN18 & 10:47:20.57 $+$59:10:40.9 & 10:47:20.54 $+$59:10:43.4 & \phantom{0}8.1 $\pm$ 1.1 & \phantom{0}7.3 $\pm$ 1.3 & \phantom{0}7.0 $\pm$ 0.8 & \phantom{0}$6.9^{\scriptscriptstyle +0.7}_{\scriptscriptstyle -0.7}$ \\ 
\hline
\end{tabular}
\end{threeparttable}
\end{table*}

\begin{table*}
\caption{SMA sample for the EGS field, ordered by decreasing deboosted SCUBA-2 flux density. All observations are from this work.}
\label{table5}
\begin{threeparttable}
\begin{tabular}{lccccccc|}
\hline                                 
Source & RA/Dec SCUBA-2 & RA/Dec SMA & $S^{\rm obs}_{\rm S2}$ [mJy] & $S_{\rm S2}$ [mJy] & $S^{\rm obs}_{\rm SMA}$ [mJy] &  $S_{\rm SMA}$ [mJy] \\
{} & (J2000) & (J2000) & {} & {} & {} & {} \\
 \hline
 EGS01 & 14:19:51.56 $+$53:00:44.8 &  14:19:51.33 $+$53:00:46.4 & 16.3 $\pm$ 1.2 & 16.3 $\pm$ 1.4 & 13.2 $\pm$ 0.9 & $12.9^{\scriptscriptstyle +0.9}_{\scriptscriptstyle -0.8}$ \\ 
 EGS02 & 14:15:57.62 $+$52:07:11.1 &  14:15:57.53 $+$52:07:12.7 & 12.7 $\pm$ 1.3 & 12.1 $\pm$ 1.2 & 13.8 $\pm$ 1.4 & $12.9^{\scriptscriptstyle +1.5}_{\scriptscriptstyle -1.2}$ \\ 
 EGS03 & 14:15:47.46 $+$52:13:47.2 &  14:15:47.09 $+$52:13:48.6 & 10.8 $\pm$ 1.0 & 10.5 $\pm$ 1.1 & 16.4 $\pm$ 2.8 & $12.9^{\scriptscriptstyle +2.9}_{\scriptscriptstyle -2.4}$ \\ 
 EGS05 & 14:19:20.35 $+$52:56:08.9 &  14:19:20.08 $+$52:56:09.1 & 10.7 $\pm$ 1.0 & 10.1 $\pm$ 1.4 & 20.0 $\pm$ 0.9 & $19.7^{\scriptscriptstyle +1.0}_{\scriptscriptstyle -0.8}$ \\ 
 EGS06 & 14:17:40.55 $+$52:29:04.7 &  14:17:40.34 $+$52:29:06.7 & 10.0 $\pm$ 1.0 & \phantom{0}9.8 $\pm$ 2.3 & \phantom{0}9.8 $\pm$ 2.0 & \phantom{0}$8.9^{\scriptscriptstyle +1.7}_{\scriptscriptstyle -1.8}$ \\ 
 EGS08 & 14:19:00.37 $+$52:49:45.3 &  14:19:00.24 $+$52:49:48.3 & 10.4 $\pm$ 1.1 & \phantom{0}9.8 $\pm$ 1.5 & \phantom{0}8.6 $\pm$ 1.5 & \phantom{0}$8.1^{\scriptscriptstyle +1.5}_{\scriptscriptstyle -1.4}$ \\ 
 EGS04 & 14:19:14.54 $+$53:00:33.6 &  14:19:14.32 $+$53:00:33.8 & 10.5 $\pm$ 1.4 & \phantom{0}9.3 $\pm$ 1.6 & 11.1 $\pm$ 1.5 & $10.5^{\scriptscriptstyle +1.2}_{\scriptscriptstyle -1.6}$ \\ 
 EGS10 & 14:17:44.09 $+$52:21:22.4 &  14:17:43.38 $+$52:21:21.7 & 10.2 $\pm$ 1.5 & \phantom{0}9.2 $\pm$ 2.3 & \phantom{0}8.3 $\pm$ 1.6 & \phantom{0}$7.7^{\scriptscriptstyle +1.6}_{\scriptscriptstyle -1.5}$ \\ 
 EGS11 & 14:17:41.73 $+$52:22:04.6 &  14:17:41.41 $+$52:22:07.9 & \phantom{0}9.8 $\pm$ 1.4 & \phantom{0}9.2 $\pm$ 1.4 & \phantom{0}7.2 $\pm$ 1.5 & \phantom{0}$6.7^{+1.6}_{-1.3}$ \\ 
\hline
 EGS07$^{\mathrm{i}}$ & 14:18:22.09 $+$52:54:01.0 &  14:18:22.04 $+$52:54:02.0 & 10.6 $\pm$ 1.6 & {} & \phantom{0}7.7 $\pm$ 1.5 & {} \\ 
 EGS09$^{\mathrm{i}}$ & 14:20:52.38 $+$52:54:02.0 &  14:20:52.55 $+$52:54:00.3 & 10.5 $\pm$ 1.6 & {} & \phantom{0}6.1 $\pm$ 1.4 & {} \\ 
 \hline
\end{tabular}
\begin{tablenotes}
\small
\item $^{\mathrm{i}}$ Source is found in the S2CLS maps but outside the area defining the S2CLS catalogue, and hence not used in our analysis. 
\end{tablenotes}
\end{threeparttable}
\end{table*}

\section{SMA imaging}

Here we provide the SMA images obtained as part of our survey. Flux density contours of the primary beam-corrected, cleaned images are shown as 2, 4 and 6 times the rms of each image. Note that these contours do not represent the actual noise used in the analysis since we used the dirty images to extract flux densities. The contours are shown over existing {\it Spitzer\/}-IRAC 3.6$\mu$m and VLA 1.4\,GHz data, when available, as well as over the parent SCUBA-2 850$\mu$m data. 

\begin{figure*}
\includegraphics[width=\textwidth]{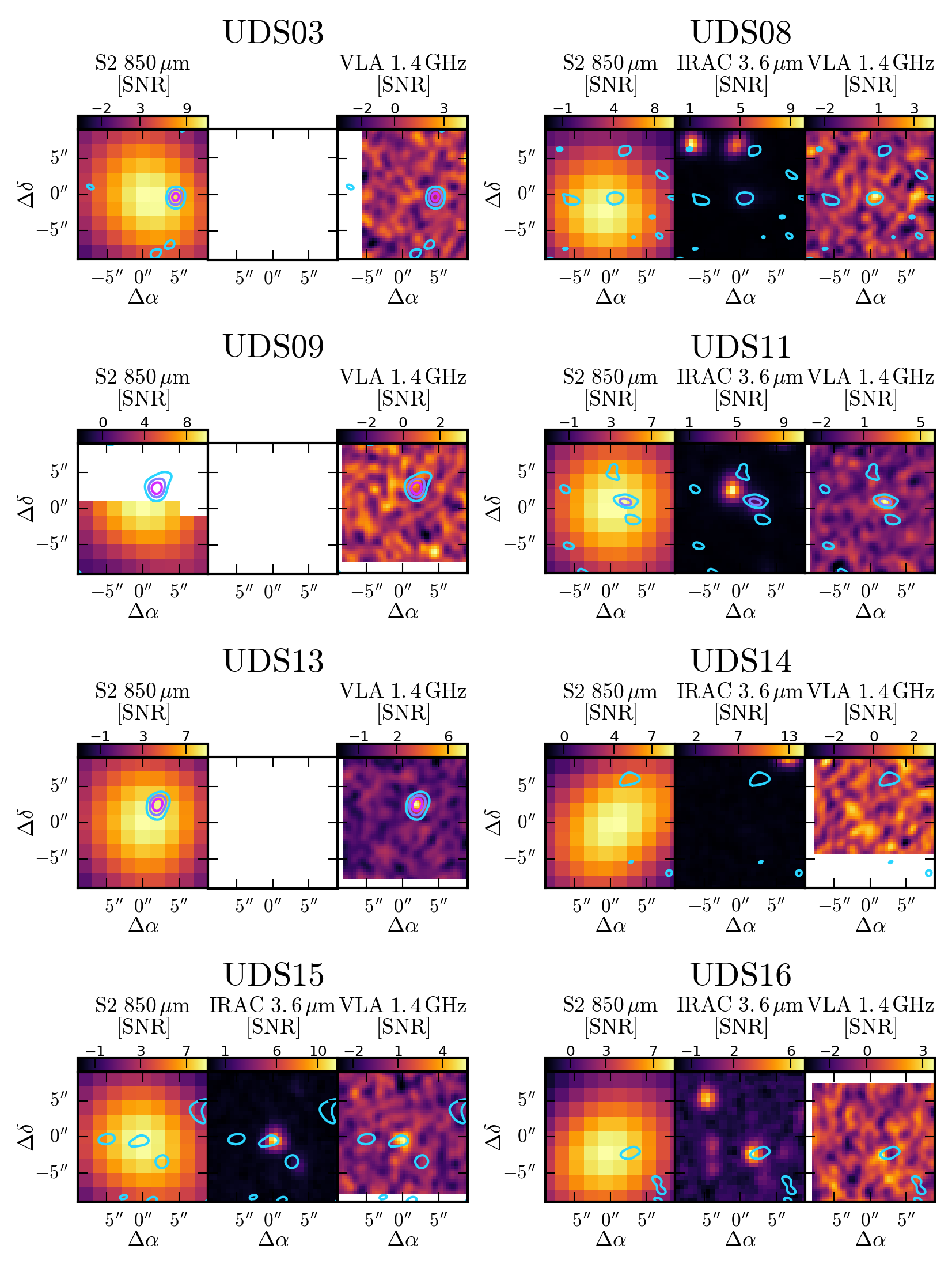}
\caption{Multiwavelength cut-outs of our primary beam-corrected, cleaned images with existing {\it Spitzer\/}-IRAC 3.6$\mu$m and VLA 1.4\,GHz imaging, when available. We show SMA flux contours of 2, 4 and 6 times the rms of each image overlaid over the IR and radio data, plus the parent SCUBA-2 850$\mu$m data.}
\label{cutouts}
\end{figure*}

\renewcommand{\thefigure}{B\arabic{figure} (Cont.)}
\addtocounter{figure}{-1}
\begin{figure*}
\includegraphics[width=\textwidth]{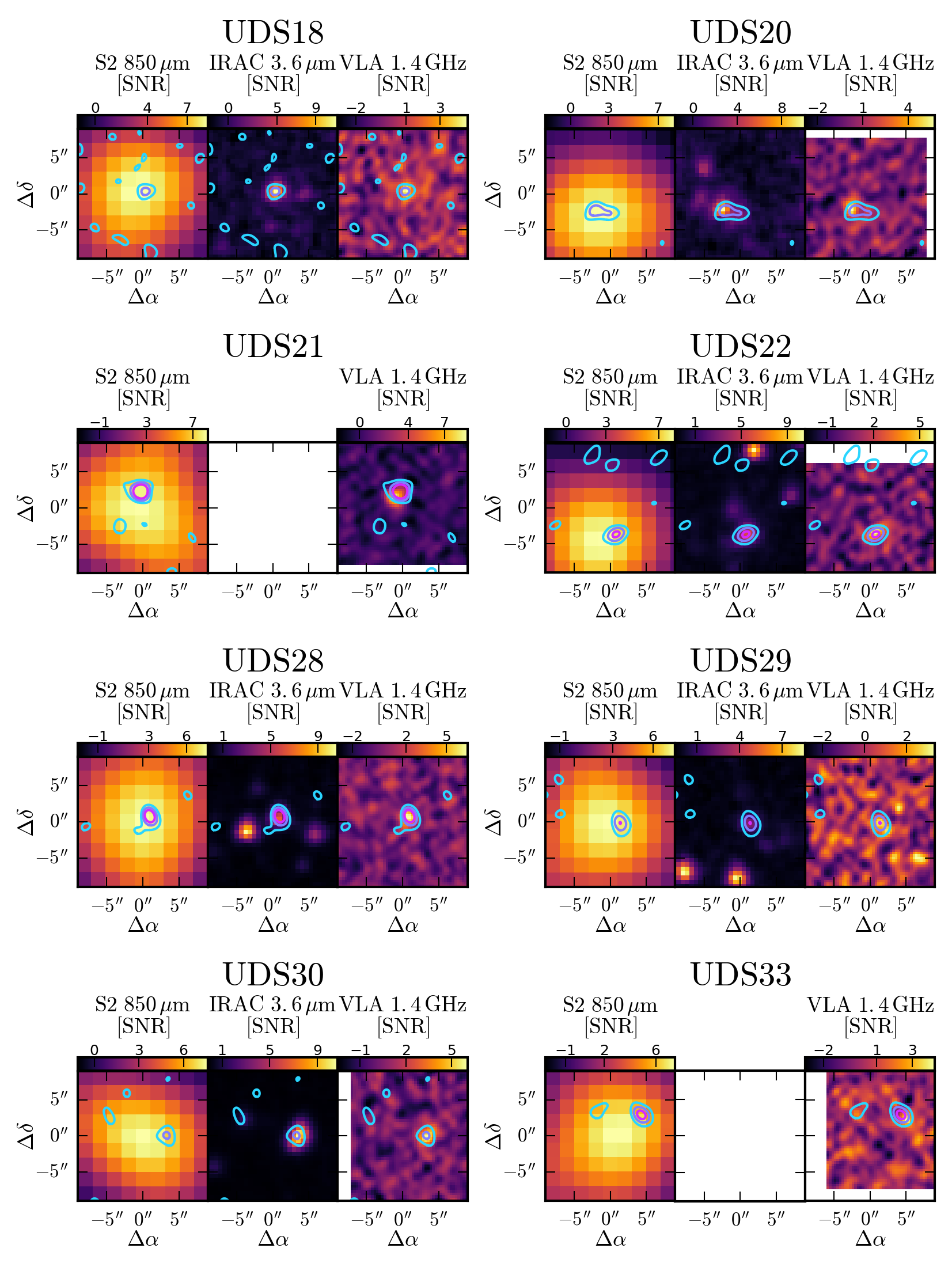}
\caption{}
\end{figure*}
\renewcommand{\thefigure}{\arabic{figure}}

\renewcommand{\thefigure}{B\arabic{figure} (Cont.)}
\addtocounter{figure}{-1}
\begin{figure*}
\includegraphics[width=\textwidth]{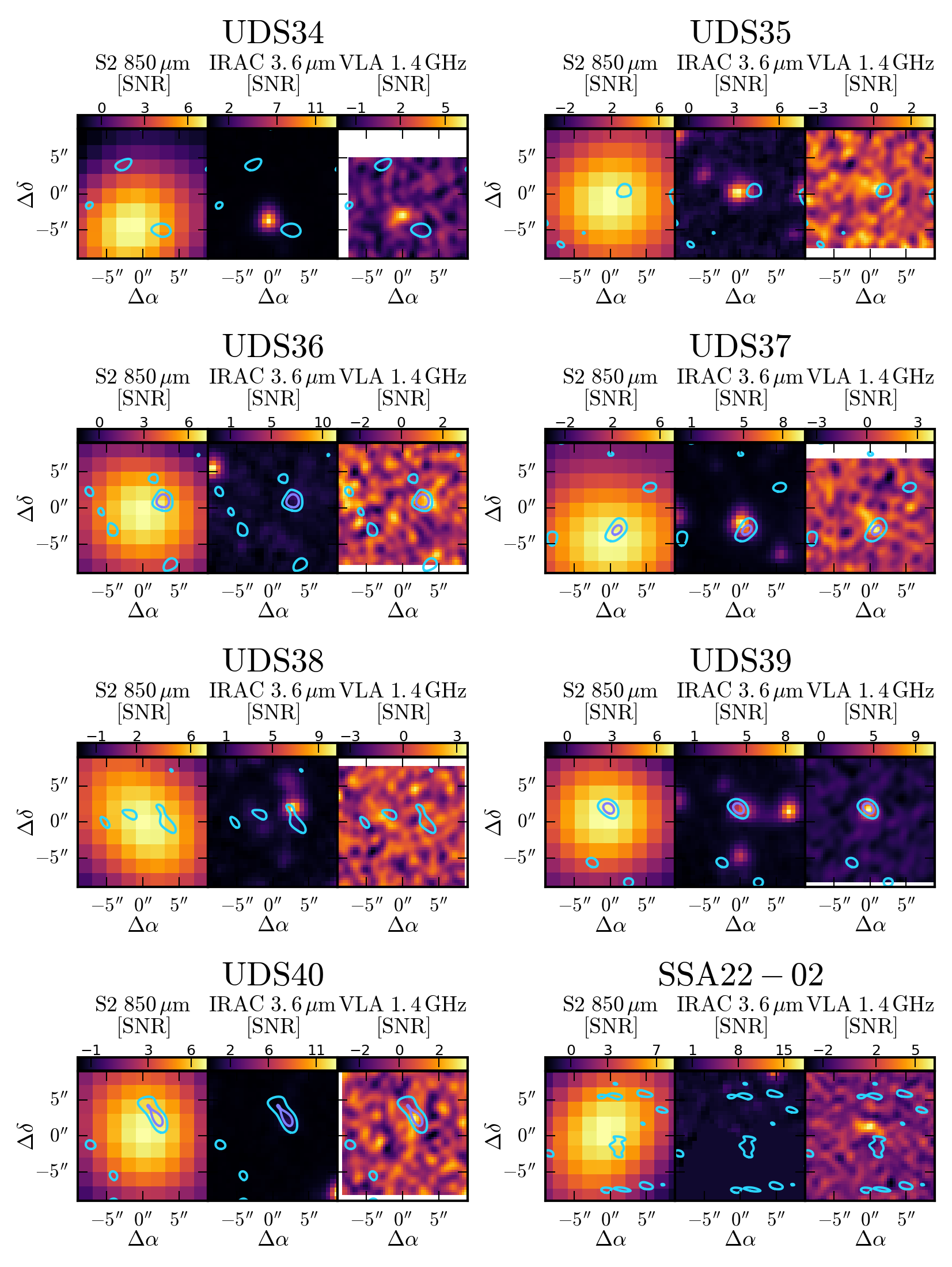}
\caption{}
\end{figure*}
\renewcommand{\thefigure}{\arabic{figure}}

\renewcommand{\thefigure}{B\arabic{figure} (Cont.)}
\addtocounter{figure}{-1}
\begin{figure*}
\includegraphics[width=\textwidth]{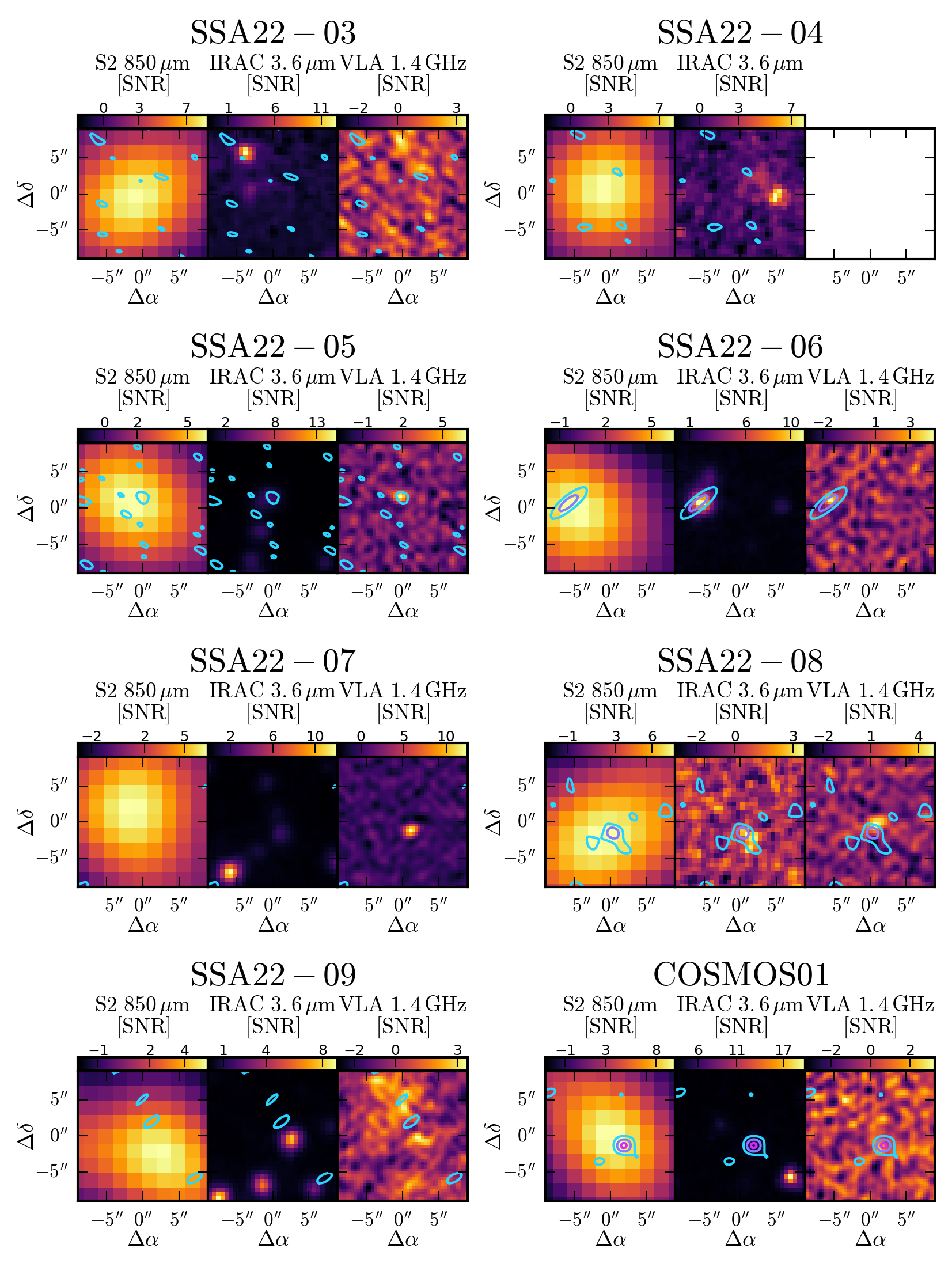}
\caption{}
\end{figure*}
\renewcommand{\thefigure}{\arabic{figure}}

\renewcommand{\thefigure}{B\arabic{figure} (Cont.)}
\addtocounter{figure}{-1}
\begin{figure*}
\includegraphics[width=\textwidth]{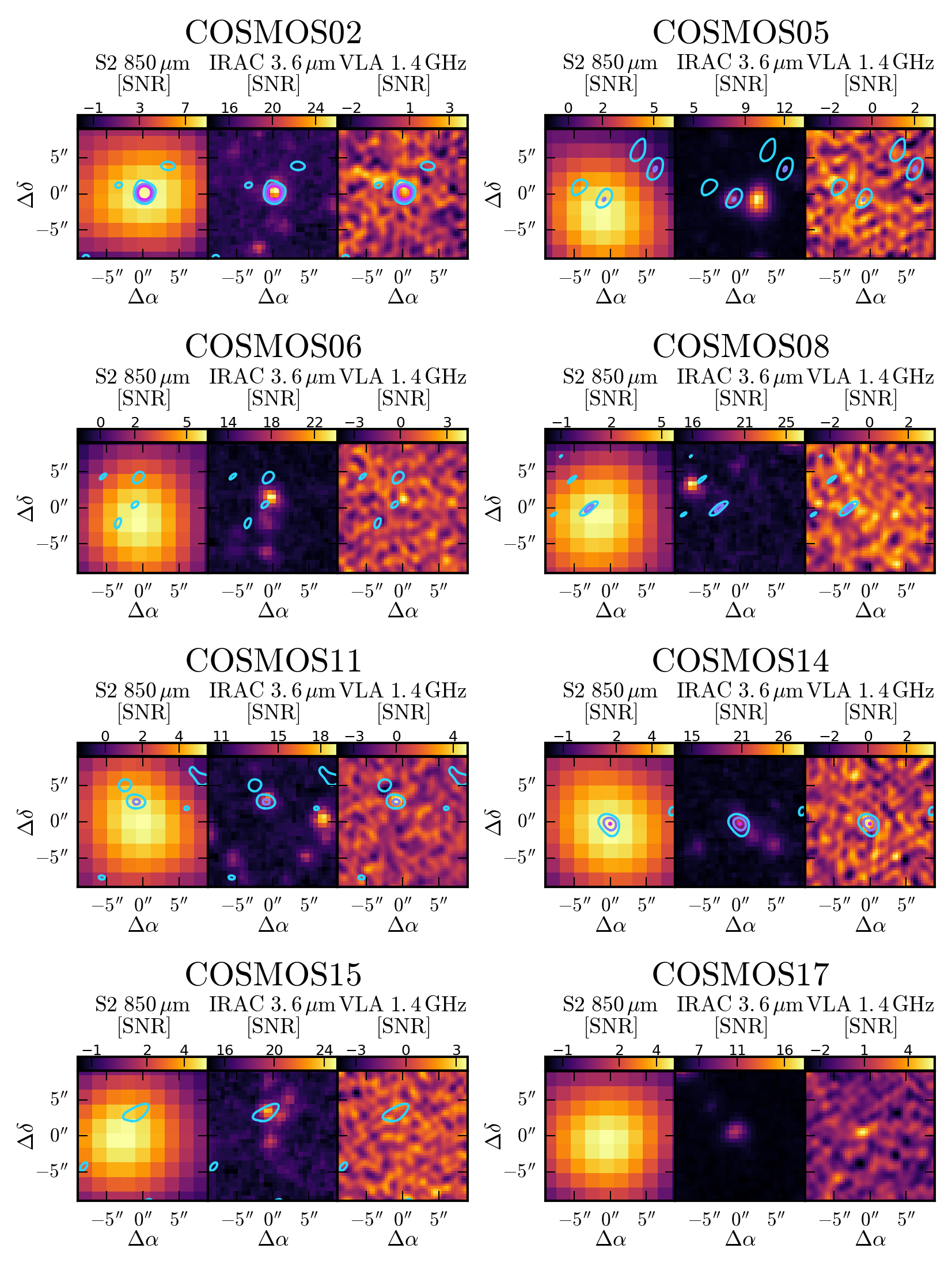}
\caption{}
\end{figure*}
\renewcommand{\thefigure}{\arabic{figure}}

\renewcommand{\thefigure}{B\arabic{figure} (Cont.)}
\addtocounter{figure}{-1}
\begin{figure*}
\includegraphics[width=\textwidth]{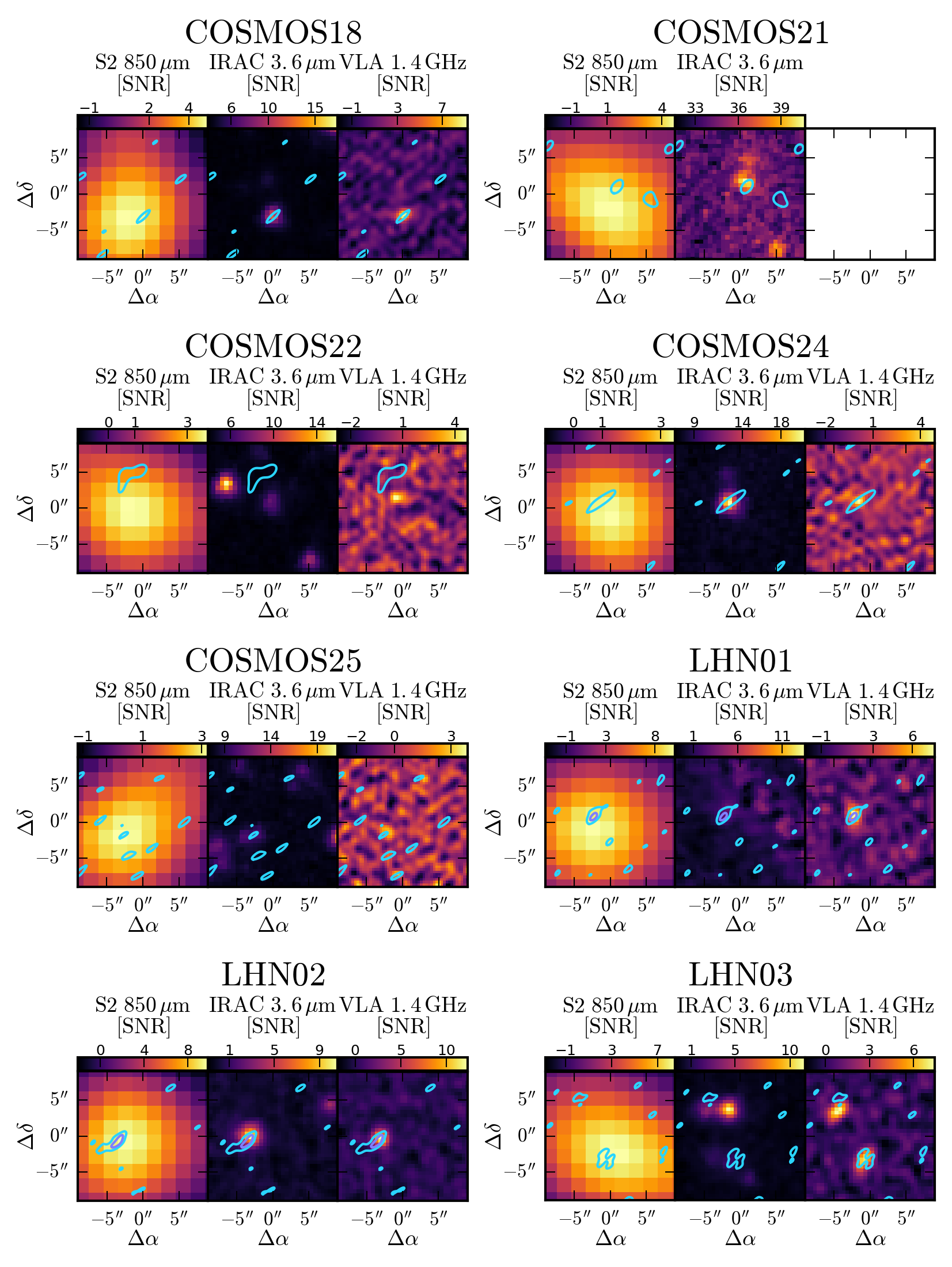}
\caption{}
\end{figure*}
\renewcommand{\thefigure}{\arabic{figure}}

\renewcommand{\thefigure}{B\arabic{figure} (Cont.)}
\addtocounter{figure}{-1}
\begin{figure*}
\includegraphics[width=\textwidth]{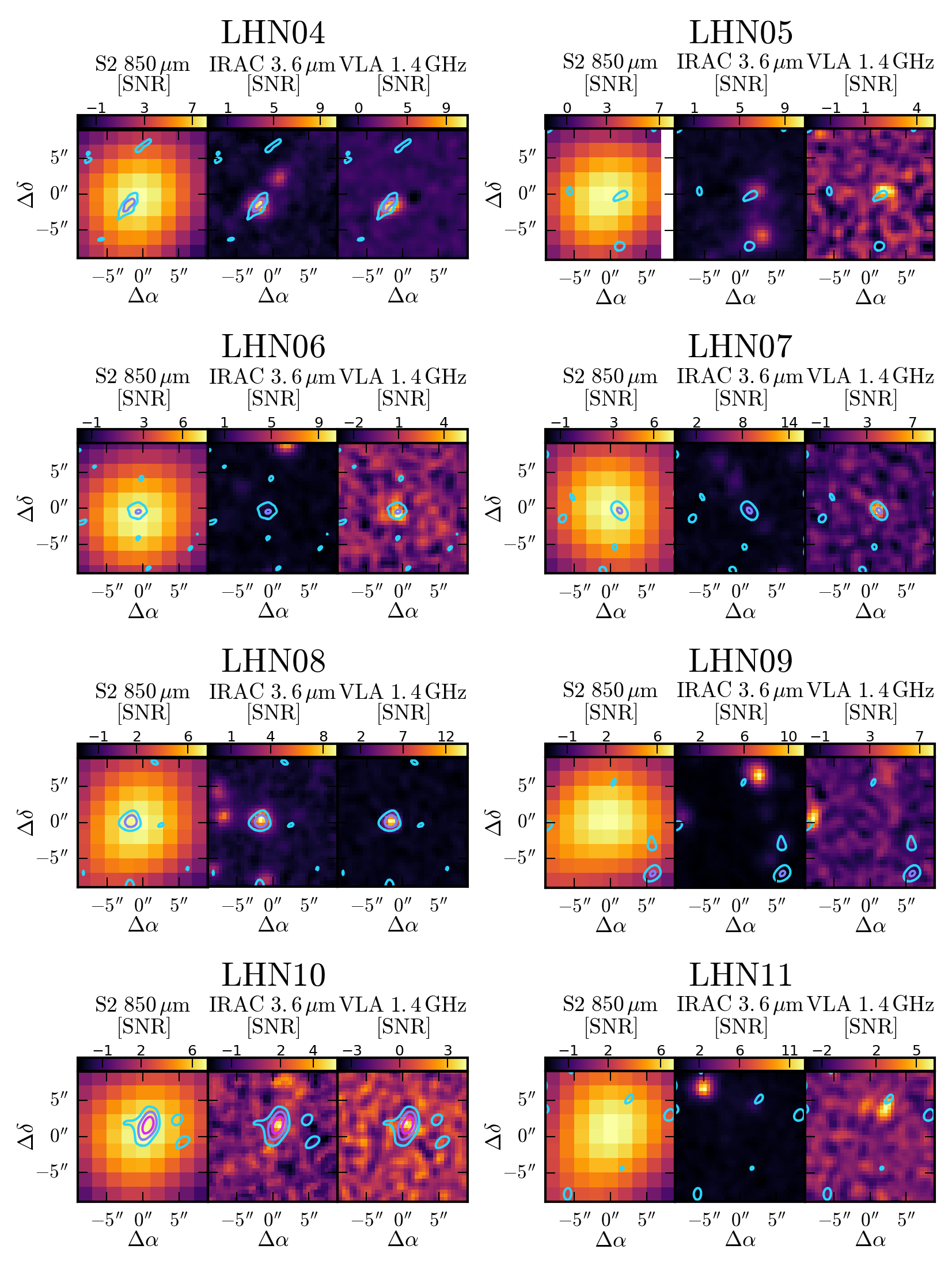}
\caption{}
\end{figure*}
\renewcommand{\thefigure}{\arabic{figure}}

\renewcommand{\thefigure}{B\arabic{figure} (Cont.)}
\addtocounter{figure}{-1}
\begin{figure*}
\includegraphics[width=\textwidth]{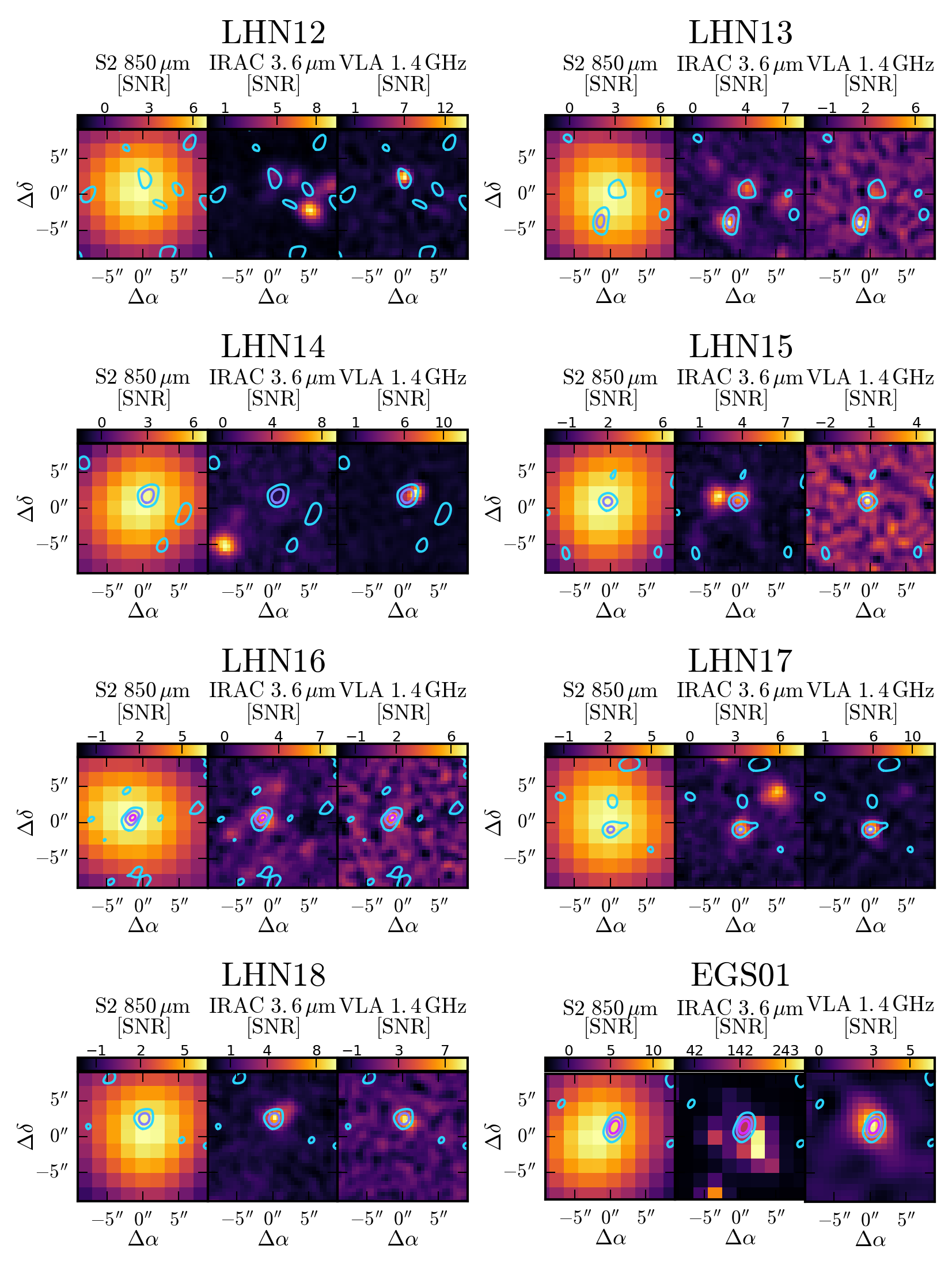}
\caption{}
\end{figure*}
\renewcommand{\thefigure}{\arabic{figure}}

\renewcommand{\thefigure}{B\arabic{figure} (Cont.)}
\addtocounter{figure}{-1}
\begin{figure*}
\includegraphics[width=\textwidth]{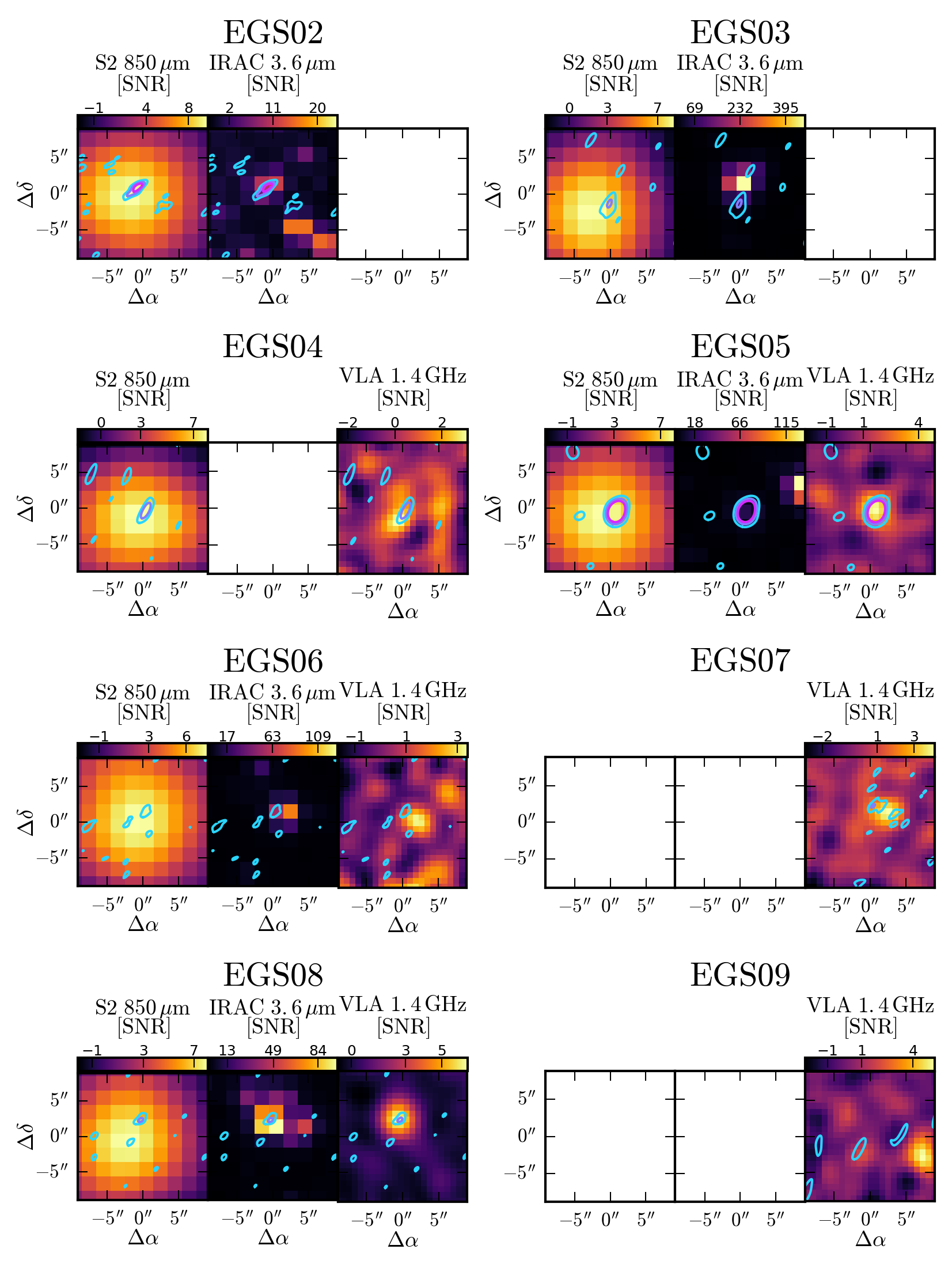}
\caption{}
\end{figure*}
\renewcommand{\thefigure}{\arabic{figure}}

\renewcommand{\thefigure}{B\arabic{figure} (Cont.)}
\addtocounter{figure}{-1}
\begin{figure*}
\includegraphics[width=\textwidth]{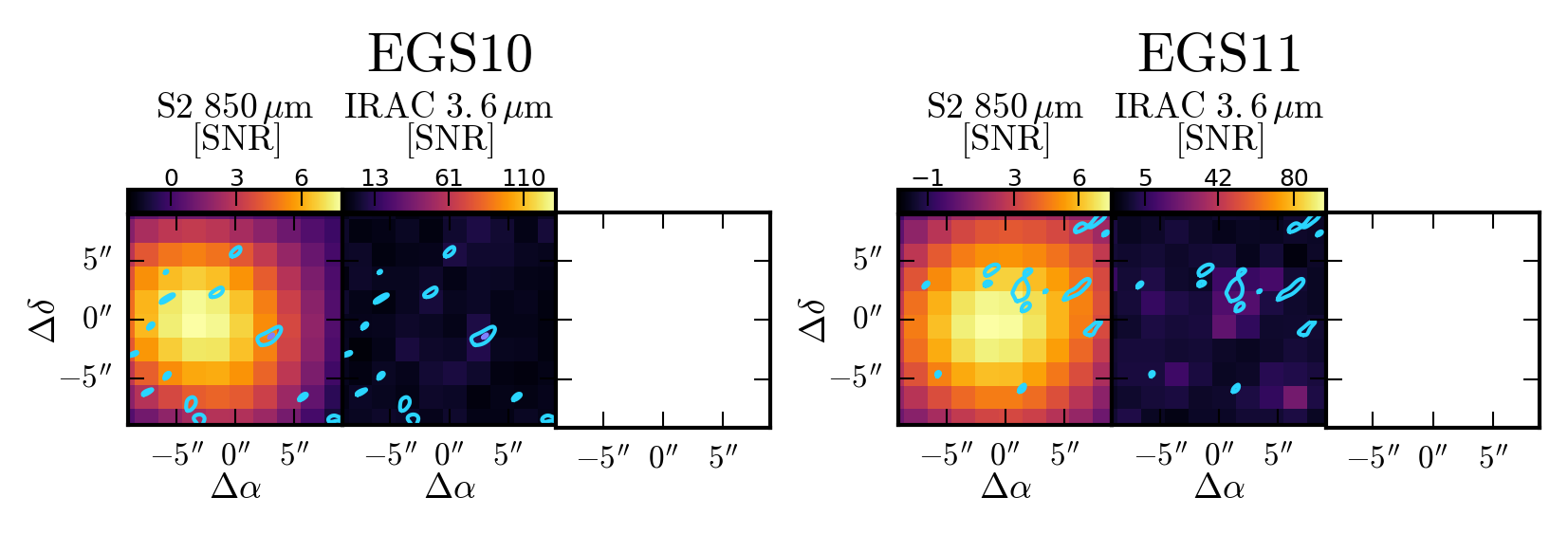}
\caption{}
\end{figure*}
\renewcommand{\thefigure}{\arabic{figure}}

\bsp	
\label{lastpage}
\end{document}